\documentclass[aps,prc,10pt,twocolumn,showpacs,floatfix]{revtex4-1}
\pdfoutput=1                 
\usepackage[utf8]{inputenc}  
\usepackage{graphicx}
\newcommand{\inputpgf}[1]{\includegraphics{#1.pdf}}

\usepackage{amsmath, amssymb}
\everymath{\displaystyle}
\renewcommand{\Re}{\operatorname{Re}}
\renewcommand{\Im}{\operatorname{Im}}
\DeclareMathOperator{\Wr}{W}
\DeclareMathOperator{\bigo}{\mathcal{O}}
\DeclareMathOperator{\fpart}{frac}
\newcommand{\D}{\mathop{}\!\mathrm{d}}
\newcommand{\E}{\mathop{}\!\mathrm{e}}
\newcommand{\I}{\mathrm{i}}
\newcommand{\anbohr}{a_{\mathrm{B}}}
\newcommand{\Ryd}{\mathrm{Ry}}
\newcommand{\fit}{\mathrm{fit}}
\newcommand{\der}[3][]{\frac{\D^{#1} #2}{\D #3^{#1}}}
\newcommand{\abs}[1]{\left|#1\right|}
\newcommand{\conj}[1]{\mkern 3mu\overline{\mkern-3mu#1}}
\newcommand{\vect}[1]{\mathbf{#1}}
\newcommand{\norm}[1]{\left\|#1\right\|}

\usepackage{hyperref}\hypersetup{pdfauthor={David Gaspard},pdftitle={Effective-range function methods for charged particle collisions},breaklinks=true}

\begin{document}
\title{Effective-range function methods for charged particle collisions} 
\author{David Gaspard}\email[E-mail:~]{dgaspard@ulb.ac.be}
\author{Jean-Marc Sparenberg}\email[E-mail:~]{jmspar@ulb.ac.be}
\affiliation{Nuclear Physics and Quantum Physics, CP229, Universit\'e libre de Bruxelles (ULB), B-1050 Brussels, Belgium}
\date{\today}

\begin{abstract}
Different versions of the effective-range function method for charged particle collisions are studied and compared.
In addition, a novel derivation of the standard effective-range function is presented from the analysis of Coulomb wave functions in the complex plane of the energy.
The recently proposed effective-range function denoted as $\Delta_\ell$ [\href{https://dx.doi.org/10.1103/PhysRevC.96.034601}{Phys. Rev. C \textbf{96}, 034601 (2017)}] and an earlier variant [\href{https://dx.doi.org/10.1016/0550-3213(73)90193-4}{Hamilton \textit{et al.}, Nucl. Phys. B \textbf{60}, 443 (1973)}] are related to the standard function.
The potential interest of $\Delta_\ell$ for the study of low-energy cross sections and weakly bound states is discussed in the framework of the proton-proton ${}^1S_0$ collision.
The resonant state of the proton-proton collision is successfully computed from the extrapolation of $\Delta_\ell$ instead of the standard function.  It is shown that interpolating $\Delta_\ell$ can lead to useful extrapolation to negative energies, provided scattering data are known below one nuclear Rydberg energy ($12.5~\mathrm{keV}$ for the proton-proton system).
This property is due to the connection between $\Delta_\ell$ and the effective-range function by Hamilton \textit{et al.} that is discussed in detail.
Nevertheless, such extrapolations to negative energies should be used with caution because $\Delta_\ell$ is not analytic at zero energy.
The expected analytic properties of the main functions are verified in the complex energy plane by graphical color-based representations.
\end{abstract}

\maketitle

\section{Introduction\label{sec:Intro}}
In quantum collision theory, the effective-range function (ERF) method is a powerful model-independent fitting technique of low-energy phase shifts~\cite{Bethe1949, Landau1944, Blatt1948, Chew1949, Jackson1950, Joachain1979, Newton1982, Haeringen1985}.
It is very useful in nuclear collision physics when the shape of the interaction potentials is not known accurately.
This method merely consists in expanding a function of the phase shift, namely the ERF, that is analytic at zero energy and behaves as a constant at this point~\cite{Bethe1949}.
The expansion of the ERF --- also referred to as the effective-range expansion --- can be either a power series of the energy or a Padé approximant, i.e., a rational function.  In general, Padé approximants are valid on a larger domain than Taylor series~\cite{Hartt1981, Rakityansky2013, Midya2015b}.
\par The ERF method was mainly developed in the 1940s by Schwinger, Bethe~\cite{Bethe1949}, Landau~\cite{Landau1944}, and others in the framework of nucleon-nucleon collisions.
In these works, it is shown that the ERF specifically dedicated to charged particle scattering is very different from the one of neutral particle scattering because of the Coulomb interaction.
In particular, the Coulomb interaction modifies the low-energy behavior of the phase shift, involving a special analytical structure described by the digamma function~\cite{Bethe1949, Abramowitz1964, Olver2010}.
Since then, the ERF for charged particle scattering has been the subject of many developments~\cite{Cornille1962, Lambert1969, Hamilton1973, Haeringen1982, Humblet1990, Yarmu2011, Rakityansky2013, Sparenberg2017, Blokhintsev2017a, Blokhintsev2017b, Orlov2016b, Orlov2017b}.
Moreover, the method has been applied to experimental data of numerous two-body systems, such as: proton-proton~\cite{Bethe1949, Landau1944, Blatt1948, Chew1949, Jackson1950, Naisse1977, Kok1980, Hartt1981, Bergervoet1988, Mukha2010}, proton-deuteron~\cite{Chen1989, Kievsky1997, Brune2001}, or ${}^{12}\mathrm{C}+\alpha$~\cite{Humblet1976, Brune1999, Filippone1989, Sparenberg2017, Blokhintsev2017a, Blokhintsev2017b, Orlov2016b, Orlov2017b}.
\par However, this ERF also raises technical issues related to the relative magnitude of its digamma function term versus the phase-shift-dependent term.
According to recent works~\cite{Sparenberg2017, Blokhintsev2017a, Blokhintsev2017b}, it would be less appropriate to model the phase shift for heavier nuclei than protons.
In Ref.~\cite{Sparenberg2017}, it is suggested to use a reduced variant of the ERF --- that is denoted as $\Delta_\ell$ --- as a potential alternative to the standard Coulomb-modified ERF for studying the weakly bound states by extrapolation of scattering data to negative energies.
This reduced ERF method is also inspired by earlier works~\cite{Hamilton1973, Humblet1990} about the mathematical properties of the standard ERF.
\par The main purpose of this paper is to study the reduced ERF method~\cite{Sparenberg2017} and to clarify its connection to the standard ERF method.
In addition, we propose a novel derivation of the standard ERF as well as the relations between the different historical formulations.
\par We show that the reduced ERF method allows us to obtain information on resonances and weakly bound states, using the properties of the digamma function appearing in the standard ERF.
These properties are graphically verified in the complex plane of the energy $E$.
Indeed, complex plots have the advantage of revealing the analytic structures that are concealed from the real $E$-axis.
This leads to predictions on the singularities of the Coulomb phase shift.
\par Finally, we apply the reduced ERF method to the proton-proton ${}^1S_0$ collision to check the predictions of the effective-range theory.
We show that the singular nature of $\Delta_\ell$ at negative energy prevents Padé approximants from converging below $E=0$.
However, depending on the energy range covered by experimental data, it seems possible with $\Delta_\ell$ to extrapolate to negative energy up to about minus one nuclear Rydberg, using the connection between the reduced ERF and the ERF by Hamilton~\textit{et al.}~\cite{Hamilton1973}, that we denote as $\Delta^+_\ell$.
\par This paper is organized as follows: Section~\ref{sec:PureCoulomb} provides the derivation of the analytic structure of the Coulomb wave functions in the complex $E$-plane.
Thereafter, the calculation of the ERFs and the study of their properties in the complex $E$-plane are presented in Section~\ref{sec:ERFs}.
This analysis is performed by explicit calculation and is aided by graphics in the complex plane.
We focus on the properties of $\Delta_\ell$ and $\Delta^+_\ell$.
Section~\ref{sec:Application} describes the theoretical properties of the ERFs in the framework of the proton-proton ${}^1S_0$ collision.
We also study the ability of the reduced ERF $\Delta_\ell$ to extrapolate data to negative energies.
\par Throughout the text, we use the reduced Planck constant $\hbar c$, the fine-structure constant $\alpha$ and the rest mass energy of the proton $m_{\mathrm{p}}c^2$ provided by the 2014 CODATA recommended values~\cite{Mohr2016}.

\section{Pure Coulomb potential\label{sec:PureCoulomb}}
This section deals with the theoretical aspects of the non-relativistic scattering of two charged particles, especially in the low-energy limit.
The Coulomb wave functions are analyzed in the complex plane of the energy.

\subsection{Scattering wave functions in Coulomb potential\label{sec:CoulombFunctions}}
The effective-range theory of the Coulomb scattering is largely based on the analytic expression of the solutions of the Schrödinger equation in a $1/r$-potential.
Indeed, the Coulomb wave functions are involved in the very definition of the phase shift~\cite{Breit1936, Bethe1949, Landau1944, Blatt1948, Chew1949, Jackson1950, Messiah1961, Gottfried1966, Joachain1979, Haeringen1985, Newton1982, Humblet1990, Rakityansky2013}.  \par We consider two spinless particles of negligible radius, respective masses $m_1$ and $m_2$, and charges $Z_1e$ and $Z_2e$.
If $\vect{r}$ denotes the relative position between the particles and $E = \hbar^2k^2/2m$ the energy in the center-of-mass frame, the Schrödinger equation writes~\cite{Messiah1961, Gottfried1966}
\begin{equation}\label{eq:PureCoulombSchrodinger}
-\frac{\hbar^2}{2m}\nabla^2\Psi(\vect{r}) + \frac{Z_1Z_2e^2}{4\pi\epsilon_0\,r}\Psi(\vect{r}) = \frac{\hbar^2k^2}{2m}\Psi(\vect{r})  \:,
\end{equation}
where $m = m_1m_2/(m_1 + m_2)$ is the reduced mass of the two-body system.
Since the Coulomb potential is isotropic --- it only depends on $r=\norm{\vect{r}}$ ---, the angular momentum commutes with the Hamiltonian.
If, in addition, we focus on a partial wave of specific angular momentum, the wave function $\Psi(\vect{r})$ splits into an angular part given by spherical harmonics $Y_{\ell m}(\theta,\phi)$ and a radial part $u_{k\ell}(r)$ to be determined~\cite{Messiah1961, Joachain1979},
\begin{equation}\label{eq:PartialWave}
\Psi_{k\ell m}(\vect{r}) = \frac{u_{k\ell}(r)}{r} Y_{\ell m}(\theta,\phi)  \:.
\end{equation}
Replacing~\eqref{eq:PartialWave} in~\eqref{eq:PureCoulombSchrodinger} and using the reduced radial coordinate $x=kr$, the Schrödinger equation~\eqref{eq:PureCoulombSchrodinger} for $u = u_{k\ell}(r)$ becomes
\begin{equation}\label{eq:CoulombWaveEq}
-\der[2]{u}{x} + \left[\frac{\ell(\ell+1)}{x^2} + \frac{2\eta}{x} - 1\right] u = 0  \:,
\end{equation}
also known as the Coulomb wave equation~\cite{Bateman1953, Abramowitz1964, Olver2010}.
The strength of the Coulomb interaction is determined by the dimensionless Sommerfeld parameter $\eta$ defined as
\begin{equation}\label{eq:SommerfeldEta}
\eta = \frac{\alpha Z_1Z_2 mc^2}{\hbar c\,k} = \frac{1}{\anbohr k}  \:,  \end{equation}
where $\anbohr$ stands for the nuclear Bohr radius (in unit length)
\begin{equation}
\anbohr = \frac{\hbar c}{\alpha Z_1Z_2 mc^2}  \:.
\end{equation}
The nuclear Bohr radius of a two-proton system is $\anbohr = 57.64~\mathrm{fm}$, which is significantly larger than the one-femtometer charge radius of the proton.
Such a large Bohr radius is due to the relatively small charge and mass of the proton compared to heavier ions.
For instance, the nuclear Bohr radius of a ${}^{12}\mathrm{C} + \alpha$ system is barely $0.806~\mathrm{fm}$, i.e., nearly a hundred times smaller.
This disparity for proton-proton scattering will play a key role in Sec.~\ref{sec:Application}~\cite{Bethe1949}.
\par We also define the nuclear Rydberg energy as
\begin{equation}
1~\Ryd = \frac{\hbar^2}{2m\anbohr^2} = \frac{1}{2}(\alpha Z_1Z_2)^2 mc^2  \:,
\end{equation}
which equals $12.49~\mathrm{keV}$ for a two-proton system and $10.72~\mathrm{MeV}$ for $^{12}\mathrm{C}+\alpha$.
\par It is well known in the literature that Eq.~\eqref{eq:CoulombWaveEq} is solved by the Coulomb wave functions~\cite{Breit1936, Bethe1949, Landau1944, Blatt1948, Chew1949, Jackson1950, Bateman1953, Messiah1961, Gottfried1966, Joachain1979, Haeringen1985, Newton1982, Abramowitz1964, Olver2010, Humblet1984, Yost1936}.
In this paper, we focus on two useful couples of linearly independent solutions of Eq.~\eqref{eq:CoulombWaveEq}: $\{F_{\eta\ell}(x), G_{\eta\ell}(x)\}$ and $\{H^+_{\eta\ell}(x), H^-_{\eta\ell}(x)\}$.  The first couple consists of the regular Coulomb function $F_{\eta\ell}(x)$ and the irregular Coulomb function $G_{\eta\ell}(x)$.
These functions are so called because of their behavior near the origin ($x=0$): the former goes like $x^{\ell+1}$ and the latter like $x^{-\ell}$ for $x\rightarrow 0$~\cite{Abramowitz1964, Olver2010}.
\par The regular Coulomb function $F_{\eta\ell}(x)$ is defined from the confluent hypergeometric function $_1F_1(a,b,z)$, also known as the Kummer function $M(a,b,z)$~\cite{Kummer1837, Abramowitz1964, Olver2010}
\begin{equation}\label{eq:KummerM}\begin{split}
M(a,b,z) & = 1 + \frac{a}{b}\frac{z}{1!} + \frac{a(a+1)}{b(b+1)}\frac{z^2}{2!} + \ldots   \\
 & = \sum_{n=0}^\infty \frac{(a)_n}{(b)_n}\frac{z^n}{n!}  \:,
\end{split}\end{equation}
where $(a)_n = a(a+1)\cdots(a+n-1) = \Gamma(a+n)/\Gamma(a)$ denotes the Pochammer symbol.
The regular Coulomb function reads~\cite{Abramowitz1964, Olver2010}
\begin{equation}\label{eq:CoulombF}
F_{\eta\ell}(x) = C_{\eta\ell} \,x^{\ell+1} \E^{\I x} M(\ell+1+\I\eta, 2\ell+2, -2\I x)  \:.
\end{equation}
It should be noted that, since $k$ appears in $\eta=1/\anbohr k$ in addition to $x=kr$, the wave number $k$ does not only act on the radial scale of the Coulomb wave functions through $x$, but it also affects the wave oscillations through $\eta$, especially for $x\lesssim 1$.
\par In the definition~\eqref{eq:CoulombF}, the normalization coefficient $C_{\eta\ell}$ ensures the far-field behavior
\begin{equation}\label{eq:FarFieldCoulombF}
F_{\eta\ell}(x) \xrightarrow{x\rightarrow\infty} \sin\left(x - \ell\frac{\pi}{2} - \eta\ln(2x) + \sigma_{\eta\ell}\right)  \:,
\end{equation}
where $\sigma_{\eta\ell}$ is the pure Coulomb phase shift~\cite{Abramowitz1964, Olver2010, Yost1936, Humblet1984}
\begin{equation}\label{eq:CoulombPhase}
\sigma_{\eta\ell} = \arg\Gamma(\ell+1+\I\eta)  \:.  \end{equation}
The coefficient $C_{\eta\ell}$ turns out to be energy-dependent~\cite{Abramowitz1964, Olver2010},
\begin{equation}\label{eq:CoulombC1}
C_{\eta\ell} = \frac{2^\ell\abs{\Gamma(\ell+1+\I\eta)}}{(2\ell+1)!\E^{\eta\pi/2}}  \quad\textrm{for}~\eta\in\mathbb{R}  \:.
\end{equation}
Here, we have to highlight that Eq.~\eqref{eq:CoulombC1} is not analytic anywhere in the complex $k$-plane because of the absolute value.
The analytic continuation of $C_{\eta\ell}$ to the complex $k$-plane is obtained by replacing $\abs{\Gamma(\ell+1+\I\eta)}$ by $[\Gamma(\ell+1+\I\eta)\Gamma(\ell+1-\I\eta)]^{1/2}$ as shown in Refs.~\cite{Lambert1969, Humblet1984} and then rewriting the product by means of Euler's reflection formula~\cite{Abramowitz1964, Olver2010}
\begin{equation}
\Gamma(1-z)\,\Gamma(1+z) = \frac{\pi z}{\sin(\pi z)}  \:.
\end{equation}
The resulting expression
\begin{equation}\label{eq:CoulombC2}
C_{\eta\ell} = \frac{(2\eta)^\ell}{(2\ell+1)!} \sqrt{\frac{2\eta\pi w_{\eta\ell}}{\E^{2\eta\pi}-1}} 
\end{equation}
is analytic in the complex $k$-plane --- except for poles and branch cuts --- and reduces to Eq.~\eqref{eq:CoulombC1} at positive energy.
In Eq.~\eqref{eq:CoulombC2}, $w_{\eta\ell}$ is a polynomial of $(\anbohr k)^2$ defined by
\begin{equation}\label{eq:CoulombW}
w_{\eta\ell} = \prod_{j=0}^\ell\left(1 + \frac{j^2}{\eta^2}\right)  \:,
\end{equation}
which equals $1$ in the zero-energy limit, as well as for $\ell=0$.
The procedure is required to make the Coulomb wave functions analytic in the energy plane. This is an important ingredient of the derivation of the Coulomb-modified ERF.
\par
\begin{figure}[ht]
\inputpgf{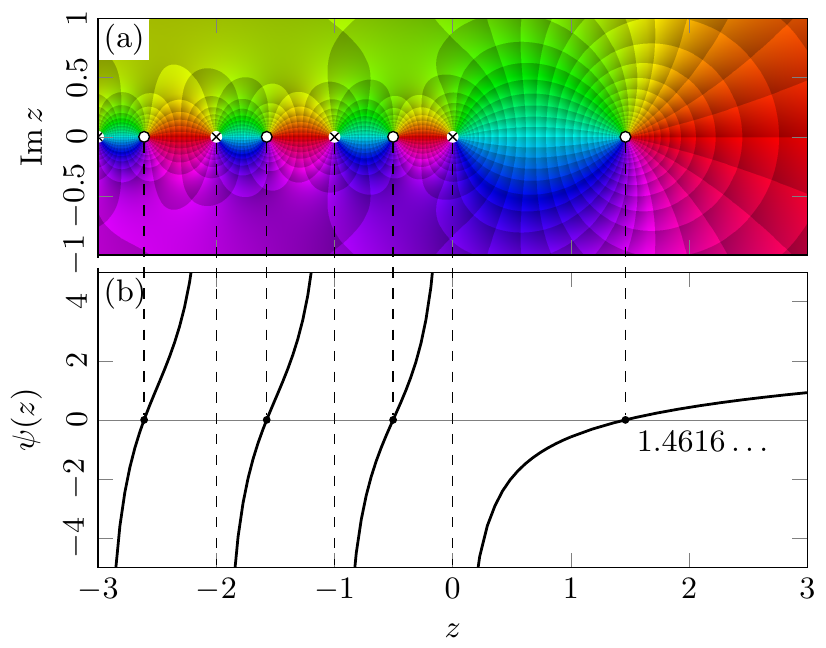}
\caption{\label{fig:plot-psi}(Color online) Representations of the digamma function $\psi$ (a) in the complex $z$-plane and (b) along the real $z$-axis.
Crosses and circles represent poles and zeroes respectively. The color legend is described in Appendix~\ref{sec:ComplexPlot}.
The poles occur when $z$ reaches a negative integer or zero.
The unique zero on the positive real axis lies at $z=1.4616\ldots$~\cite{Olver2010}.}
\end{figure}

 Before talking about the irregular Coulomb function $G_{\eta\ell}(x)$, we have to introduce the incoming and outgoing Coulomb wave functions, respectively denoted as $H^-_{\eta\ell}(x)$ and $H^+_{\eta\ell}(x)$.
They are defined in a very similar way to Eq.~\eqref{eq:CoulombF} by
\begin{equation}\label{eq:CoulombH}
H^\pm_{\eta\ell}(x) = D^\pm_{\eta\ell} \,x^{\ell+1} \E^{\pm\I x} U(\ell+1\pm\I\eta, 2\ell+2, \mp2\I x)  \:,
\end{equation}
where $U(a,b,z)$ is the confluent hypergeometric function of the second kind, also known as Tricomi's function, which is linearly independent of $M(a,b,z)$~\cite{Bateman1953, Abramowitz1964, Olver2010}.
The Tricomi function in Eq.~\eqref{eq:CoulombH} is defined by a peculiar series representation.
Because $b$ is an integer, $U(a,b,z)$ splits into two parts~\cite{Olver2010} that we call $P(a,b,z)$ and $L(a,b,z)$
\begin{equation}\label{eq:TricomiU}  U(a, b, z) = P(a,b,z) + L(a,b,z)  \quad\text{for}~b\in\mathbb{Z}^+  \:.
\end{equation}
In the following, we occasionally use the symbols $a=\ell+1+\I\eta$, $b=2\ell+2$ and $z=-2\I kr$ borrowed from confluent hypergeometric functions~\cite{Olver2010} to shorten the notations of the Coulomb wave functions.
In Eq.~\eqref{eq:TricomiU}, the first term is a polynomial of negative powers of $z$~\cite{Olver2010}
\begin{equation}\label{eq:TricomiP}
P(a,b,z) = \frac{(2\ell)!}{\Gamma(\ell+1+\I\eta)} \sum_{n=0}^{2\ell} \frac{(-\ell+\I\eta)_n}{(-2\ell)_n} \frac{z^{n-2\ell-1}}{n!}  \:,
\end{equation}
and the second one is a generalized series involving logarithmic terms in $z$~\cite{Olver2010}
\begin{equation}\label{eq:TricomiL}\begin{split}
L(a,b,z) & = \frac{(-1)^{2\ell+2}}{(2\ell+1)!\,\Gamma(-\ell+\I\eta)} \sum_{n=0}^\infty\frac{(a)_n}{(b)_n}\frac{z^n}{n!}   \\
 \times  & \big[\ln z + \psi(a+n) - \psi(b+n) - \psi(n+1)\big]   \:,
\end{split}\end{equation}
where $\psi(z)$ is the digamma function --- also known as the psi function --- defined as the logarithmic derivative of the gamma function
\begin{equation}\label{eq:PsiDef}
\psi(z) = \Gamma'(z)/\Gamma(z)  \:.
\end{equation}
The function $\psi(z)$ is shown in the complex $z$-plane in Fig.~\ref{fig:plot-psi}(a).
One notices the array of poles and zeroes on the negative real $z$-axis; they play an important role in the following.
\par The normalization coefficients $D^\pm_{\eta\ell}$ in Eq.~\eqref{eq:CoulombH}, that we define as
\begin{equation}\label{eq:CoulombD}
D^\pm_{\eta\ell} = \mp2\I(-1)^\ell\E^{\eta\pi} \frac{(2\ell+1)!\,C_{\eta\ell}}{\Gamma(\ell+1\mp\I\eta)}  \:,
\end{equation}
are intended to ensure the asymptotic behavior
\begin{equation}\label{eq:FarFieldCoulombH}
H^\pm_{\eta\ell}(x) \xrightarrow{x\rightarrow\infty} \exp\left\{\pm\I\left[x - \ell\frac{\pi}{2} - \eta\ln(2x) + \sigma_{\eta\ell}\right]\right\}   \:.
\end{equation}
There are other ways to define $D^\pm_{\eta\ell}$ in the literature~\cite{Yost1936, Abramowitz1964, Olver2010, Humblet1984}, but Eq.~\eqref{eq:CoulombD} has the advantage of being proportional to $C_{\eta\ell}$.
It will play an important role in Sec.~\ref{sec:CoulombStructure}.
\par Although $D^+_{\eta\ell}$ and $D^-_{\eta\ell}$, as well as $H^+_{\eta\ell}(x)$ and $H^-_{\eta\ell}(x)$, are related to each other by complex conjugation at real $k$, this no longer holds when $k$ is complex valued.
This is because the complex conjugation is not an analytic operation: $\conj{z}$ cannot be expanded in power series of $z$ for $z\in\mathbb{C}$.
However, one can resort to complex conjugation (in the sense of Ref.~\cite{Hamming1973}) to relate them for complex-valued $k$
\begin{equation}\label{eq:CoulombConj}\begin{cases}
D^-_{\eta\ell} = \conj{D^+_{\conj{\eta}\ell}}                 \:,\\
H^-_{\eta\ell}(kr) = \conj{H^+_{\conj{\eta}\ell}(\conj{k}r)}  \:,
\end{cases}\end{equation}
where $\conj{\eta} = 1/\anbohr\conj{k}$.
The relations~\eqref{eq:CoulombConj} are practically obtained by replacing everywhere $+\I$ by $-\I$, and conversely.
\par Contrary to $F_{\eta\ell}(x)$, the functions $H^\pm_{\eta\ell}(x)$ are complex valued and irregular at $x=0$ like $x^{-\ell}$.
However, the regular Coulomb function can be retrieved by subtracting $H^-_{\eta\ell}(x)$ from $H^+_{\eta\ell}(x)$
\begin{equation}
F_{\eta\ell}(x) = \frac{H^+_{\eta\ell}(x) - H^-_{\eta\ell}(x)}{2\I}  \:,
\end{equation}
which reduces for real $k$ to $\Im H^+_{\eta\ell}(x)$ and to the sine wave of Eq.~\eqref{eq:FarFieldCoulombF} consistently with Eq.~\eqref{eq:FarFieldCoulombH}.
\par Finally, the irregular Coulomb function $G_{\eta\ell}(x)$ is defined as~\cite{Yost1936, Messiah1961, Humblet1984, Olver2010}
\begin{equation}\label{eq:CoulombG}
G_{\eta\ell}(x) = \frac{H^+_{\eta\ell}(x) + H^-_{\eta\ell}(x)}{2}  \:.
\end{equation}
This definition has to be understood as the real part $\Re H^\pm_{\eta\ell}(x)$ only on the real $k$-axis.
Otherwise, when $k$ is complex --- at negative energies for instance --- Eq.~\eqref{eq:CoulombG} should be preferred, being the analytic continuation of $\Re H^\pm_{\eta\ell}(x)$.
\par From Eq.~\eqref{eq:FarFieldCoulombH}, one easily shows that the irregular function $G_{\eta\ell}(x)$ behaves asymptotically like
\begin{equation}
G_{\eta\ell}(x) \xrightarrow{x\rightarrow\infty} \cos\left(x - \ell\frac{\pi}{2} - \eta\ln(2x) + \sigma_{\eta\ell}\right)  \:.
\end{equation}
\par It should be noted that the $s$ wave ($\ell=0$) of $G_{\eta\ell}(x)$ shows a peculiarity in the low-range limit $r\rightarrow 0$.
Instead of behaving like $\bigo(1)$ as predicted by $\bigo(x^{-\ell})$, the irregular wave function $G_{\eta0}(x)$ is exceptionally dominated by a logarithmic term $\bigo(x\ln x)$ emanating from the series~\eqref{eq:TricomiL} in $U(a,b,z)$.
As will be seen later in Sec.~\ref{sec:CoulombStructure}, such logarithmic terms affect the properties of the irregular Coulomb function in the complex $k$-plane, and in this way the effective-range function.

\subsection{Analytic structure of Coulomb wave functions\label{sec:CoulombStructure}}
In this section, we propose a new derivation of the analytic structure of the Coulomb wave functions in the complex plane of the energy $E$, especially in the low-energy limit.
We show in Sec.~\ref{sec:UsualERF} how the structure of the irregular Coulomb functions $H^\pm_{\eta\ell}(x)$ and $G_{\eta\ell}(x)$ leads to the standard effective-range function.
To this end, it is crucial to study the analytic properties of the Coulomb wave functions, since they are involved in the very definition of the phase shift $\delta_\ell(E)$. \par For simplicity, we focus our analysis on Coulomb wave functions divided by the normalization factor $C_{\eta\ell}$ of Eq.~\eqref{eq:CoulombC2}.
Indeed, this factor will disappear from the calculation of the phase shift because the wave function $u_{k\ell}(r)$ is defined within a (possibly complex) factor.
\par Let us begin with the regular Coulomb function $F_{\eta\ell}(x)$.
From the definition~\eqref{eq:CoulombF} and the power series~\eqref{eq:KummerM}, one has
\begin{equation}\label{eq:CoulombFSeries}
\frac{F_{\eta\ell}(kr)}{C_{\eta\ell}(kr)^{\ell+1}\E^{\I kr}} = \sum_{n=0}^\infty \frac{(\ell+1+\I\eta)_n}{(2\ell+2)_n}\frac{(-2\I kr)^n}{n!}  \:.
\end{equation}
When the wave number $k$ vanishes, $\eta = 1/\anbohr k$ tends to infinity.
Fortunately, the Pochammer symbol $(\ell+1+\I\eta)_n$ is asymptotic to $(\I\eta)^n$.
The numerator in the right-hand side of Eq.~\eqref{eq:CoulombFSeries} becomes $(2\eta kr)^n = (2r/\anbohr)^n$ which no longer depends on $k$.
Therefore, the series is well defined at zero energy and behaves as a constant in the neighborhood of $k=0$.
\par The analytic structure of the irregular Coulomb functions $H_{\eta\ell}^\pm(x)$ and $G_{\eta\ell}(x)$ is less obvious than for $F_{\eta\ell}(x)$ but it has important consequences on the ERF.
For convenience, we begin the study with the outgoing Coulomb wave function $H^+_{\eta\ell}(x)$ instead of $G_{\eta\ell}(x)$.
Given Eqs.~\eqref{eq:CoulombConj} and~\eqref{eq:CoulombG}, all the equations below for $H^+_{\eta\ell}(x)$ will impact those for $G_{\eta\ell}(x)$.
\par Let us begin the analysis of $H^+_{\eta\ell}(x)$ in the $k$-plane with $U(a,b,z)$ from Eq.~\eqref{eq:TricomiU}.
The two functions $P(a,b,z)$ and $L(a,b,z)$ from Eqs.~\eqref{eq:TricomiP} and~\eqref{eq:TricomiL} are singular in the neighborhood of $k=0$, and one needs to regularize both of them in the limit $k\rightarrow 0$.
\par First, we consider the finite sum $P(a,b,z)$ from Eq.~\eqref{eq:TricomiP}.
This function is singular at zero energy on the one hand because of the essential singularity at $k=0$ in $\Gamma(\ell+1+\I\eta)^{-1}$ and on the other hand because each term in the series behaves like $k^{-(2\ell+1)}$ as $k\rightarrow 0$.
One way to circumvent this issue is to define the regularized function
\begin{equation}\label{eq:CoulombP}
P^+_{\eta\ell}(x) = \frac{(2\ell+1)!}{(\I\eta)^{2\ell+1}} \Gamma(\ell+1+\I\eta) \,P(a,b,z)  \:,
\end{equation}
which is holomorphic for $k\in\mathbb{C}$ due to the compensation of all the singularities. \par Regarding the series $L(a,b,z)$ of Eq.~\eqref{eq:TricomiL}, three kinds of singularities have to be considered while regularizing~\cite{Cornille1962, Lambert1969}:
\begin{enumerate}
\item the essential singularity of $\Gamma(-\ell+\I\eta)^{-1}$ at $k=0$,
\item the branch cut of the principal-valued logarithm $\ln z$ in the series, and
\item the array of poles of the digamma function $\psi(a+n)$ when $a+n\in\{0, -1, -2, \ldots\}$ leading to an accumulation point at $k=0$.
\end{enumerate}
First, the gamma function $\Gamma(-\ell+\I\eta)^{-1}$ can be compensated in the same way as with $P(a,b,z)$.
Secondly, the energy dependence of $\ln z$ can be separated from the radial part as
\begin{equation}\label{eq:LogSeparation}
\ln(z) = \ln(-2\I kr) = \ln(2r/\anbohr) - \ln(\I\eta)  \:.
\end{equation}
It should be noted that the above decomposition assumes that the two particles are repelling each other ($\anbohr>0$).
When it is not true, one can choose $\ln(-2\I kr) = \ln(-2r/\anbohr) - \ln(-\I\eta)$ instead without other significant difference in the calculation.
\par Finally, the most difficult part of the regularization of $H^+_{\eta\ell}(x)$ involves the digamma function $\psi(a+n)$ in the series~\eqref{eq:TricomiL}.
The digamma function $\psi(a+n)$ has infinitely many poles on the imaginary $k$-axis with an accumulation point at zero energy ($k=0$)
\begin{equation}
k \in \left\{\frac{-\I}{(\ell+n+1)\anbohr}, \frac{-\I}{(\ell+n+2)\anbohr}, \ldots, \rightarrow -0\,\I\right\}  \:.
\end{equation}
Such a structure is not compensated by the zeroes from $(a)_n$.
Therefore, one has to separate the digamma function $\psi(a+n)$ from the series~\eqref{eq:TricomiL}.
For this purpose, one uses the property~\cite{Abramowitz1964, Olver2010}
\begin{equation}\label{eq:PsiSeparation}
\psi(a+n) = \psi(a) + \sum_{s=0}^{n-1}\frac{1}{a+s}  \:,
\end{equation}
to extract from $\psi(a+n)$ the function $\psi(a)$ independent of the summation index $n$, but still with the $k$-dependence.
\par From this point on, we exploit the close similarity between the series~\eqref{eq:TricomiL} and the Kummer function~\eqref{eq:KummerM}.
Indeed, we guess that the index-independent part will contribute to a Kummer function $M(a,b,z)$, that is holomorphic for $k\in\mathbb{C}$ as shown before in~\eqref{eq:CoulombFSeries}.
Using Eqs.~\eqref{eq:LogSeparation} and~\eqref{eq:PsiSeparation}, one gets the analytic decomposition
\begin{equation}\label{eq:TricomiLStructure}
L(a,b,z) = \frac{g_\ell^+(\eta)\,M(a,b,z) + \sum_{n=0}^\infty c_n\frac{(a)_n}{(b)_n}\frac{z^n}{n!}}{(2\ell+1)!\,\Gamma(-\ell+\I\eta)}  \:,
\end{equation}
where $g^+_\ell(\eta)$ is given by  \begin{equation}\label{eq:BetheGpm}
g^\pm_\ell(\eta) = \psi(\ell + 1 \pm\I\eta) - \ln(\pm\I\eta)  \:.
\end{equation}
\par The coefficients $c_n$ of the series in Eq.~\eqref{eq:TricomiLStructure} contain the terms
\begin{equation}\label{eq:TricomiSeriesCoeff}
c_n = \ln\left(\frac{2r}{\anbohr}\right) + \sum_{s=0}^{n-1}\frac{1}{a+s} - \psi(b+n) - \psi(n+1)  \:,
\end{equation}
which remain after the decomposition.
Other equivalent decompositions may lead to different functions $g^\pm_\ell(\eta)$ and coefficients $c_n$. The key point is to realize that the hypergeometric-like series in Eq.~\eqref{eq:TricomiLStructure},
\begin{equation}
\sum_{n=0}^\infty c_n\frac{(a)_n}{(b)_n}\frac{z^n}{n!}  \:,
\end{equation}
is holomorphic in the $k$-plane due to compensation between the zeroes of $(a)_n z^n$ and the poles of
\begin{equation}
\sum_{s=0}^{n-1} \frac{1}{a + s}  \:,
\end{equation}
in the coefficients $c_n$ of Eq.~\eqref{eq:TricomiSeriesCoeff}.
Similarly to what has been done before for $P(a,b,z)$, we define a new $k$-holomorphic function
\begin{equation}\label{eq:CoulombL}
L^+_{\eta\ell}(x) = \frac{\Gamma(\ell+1+\I\eta)}{(\I\eta)^{2\ell+1}\Gamma(-\ell+\I\eta)} \sum_{n=0}^\infty c_n\frac{(a)_n}{(b)_n}\frac{z^n}{n!}  \:.
\end{equation}
\par The prefactor in Eq.~\eqref{eq:CoulombL} is obtained by multiplying~\eqref{eq:TricomiLStructure} by the same coefficient as $P(a,b,z)$ of Eq.~\eqref{eq:CoulombP}.
The idea behind this renormalization is to keep $L^+_{\eta\ell}(x)$ on the same footing as $P^+_{\eta\ell}(x)$.
The resulting prefactor in Eq.~\eqref{eq:CoulombL} is also holomorphic in $k$ as evidenced by the corollary of the recurrence property of the gamma function~\cite{Humblet1984}
\begin{equation}
\frac{\Gamma(\ell+1+\I\eta)}{(\I\eta)^{2\ell+1}\Gamma(-\ell+\I\eta)} = w_{\eta\ell}   \:,
\end{equation}
with the polynomial $w_{\eta\ell}$ given by Eq.~\eqref{eq:CoulombW}.
\par From now on, one can rewrite the outgoing Coulomb function $H^+_{\eta\ell}(x)$ of Eq.~\eqref{eq:CoulombH} in term of the $k$-holomorphic functions $P^+_{\eta\ell}(x)$ and $L^+_{\eta\ell}(x)$ using Eqs.~\eqref{eq:TricomiU}, \eqref{eq:CoulombP}, \eqref{eq:TricomiLStructure} and~\eqref{eq:CoulombL}.
One gets the analytic decomposition
\begin{equation}\begin{split}
H^+_{\eta\ell}(x) & = \frac{D_{\eta\ell}^+ \,x^{\ell+1}\E^{\I x}}{(2\ell+1)!\,\Gamma(-\ell+\I\eta)} g_\ell^+(\eta)\,M(a,b,z)   \\
 + & \frac{(\I\eta)^{2\ell+1} D_{\eta\ell}^+ \,x^{\ell+1}\E^{\I x}}{(2\ell+1)!\,\Gamma(\ell + 1 + \I\eta)} \left[P^+_{\eta\ell}(x) + L^+_{\eta\ell}(x)\right]  \:.
\end{split}\end{equation}
The coefficients above can be further simplified using the relation~\eqref{eq:CoulombD} between the normalization factors $D^\pm_{\eta\ell}$ and $C_{\eta\ell}$.
In addition, we introduce the regularized Coulomb wave functions
\begin{equation}\label{eq:CoulombIpm}
I^\pm_{\eta\ell}(x) = C_{\eta\ell} \,x^{\ell+1}\E^{\pm\I x} \left[P^\pm_{\eta\ell}(x) + L^\pm_{\eta\ell}(x)\right]  \:,
\end{equation}
such that $I^\pm_{\eta\ell}(x)/C_{\eta\ell}k^{\ell+1}$ behaves as a constant when $k$ tends to zero.
Then, we end up with the analytic decomposition
\begin{equation}\label{eq:CoulombHStructure}
H^+_{\eta\ell}(x) = \frac{\E^{2\eta\pi}-1}{\pi}\left[g^+_\ell(\eta) F_{\eta\ell}(x) + \frac{1}{w_{\eta\ell}} I^+_{\eta\ell}(x)\right]  \:.
\end{equation}
All the singularities of $H^+_{\eta\ell}(x)/C_{\eta\ell}$ in the $k$-plane originate from $\E^{2\eta\pi}$, $g^+_\ell(\eta)$ and $1/w_{\eta\ell}$.
Therefore, Eq.~\eqref{eq:CoulombHStructure} can be understood as a kind of factorization of the singular $k$-dependence of $H^+_{\eta\ell}(x)$ from the functions $F_{\eta\ell}(x)$ and $I^+_{\eta\ell}(x)$ that are regular in $k$ except for their common normalization coefficient $C_{\eta\ell}$~\cite{Rakityansky2013}.
\par Finally, the analytic decomposition of the irregular Coulomb function $G_{\eta\ell}(x)$ can be derived from Eq.~\eqref{eq:CoulombHStructure} and the definition~\eqref{eq:CoulombG}.
Indeed, the decomposition of $H^-_{\eta\ell}(x)$ is obtained by changing the signs of $g^+_\ell(\eta)$ and $I^+_{\eta\ell}(x)$ in Eq.~\eqref{eq:CoulombHStructure}.
It is more convenient to define a new function
\begin{equation}\label{eq:BetheG}
g_{\ell}(\eta) = \frac{g^+_{\ell}(\eta) + g^-_{\ell}(\eta)}{2}  \:,
\end{equation}
in the same way as $G_{\eta\ell}(x)$.
In this paper, we refer to $g_{\ell}(\eta)$ as ``Bethe's function''~\cite{Bethe1949} although it was concurrently found by Landau~\cite{Landau1944}.
This singular function $g_{\ell}(\eta)$ is discussed in details in Subsection~\ref{sec:BetheAnalysis}.
Besides, we define the real-valued regularized Coulomb function $I_{\eta\ell}(x)$ accordingly
\begin{equation}\label{eq:CoulombI}
I_{\eta\ell}(x) = \frac{I^+_{\eta\ell}(x) + I^-_{\eta\ell}(x)}{2}  \:,
\end{equation}
which is related within a factor to similar functions found in the literature: $u_S$~\cite{Cornille1962}, $\theta_{\ell}$~\cite{Lambert1969}, $\Omega_{\ell}$~\cite{Humblet1990} or $\Psi_{\ell}$~\cite{Yost1936, Breit1936, Humblet1984}.
\par It should be noted that the functions $I^\pm_{\eta\ell}(x)$ as well as $I_{\eta\ell}(x)$ are also solutions of the Schrödinger equation~\eqref{eq:CoulombWaveEq}, because they are linear combinations of the regular and the irregular Coulomb functions.
\begin{figure}[ht]
\inputpgf{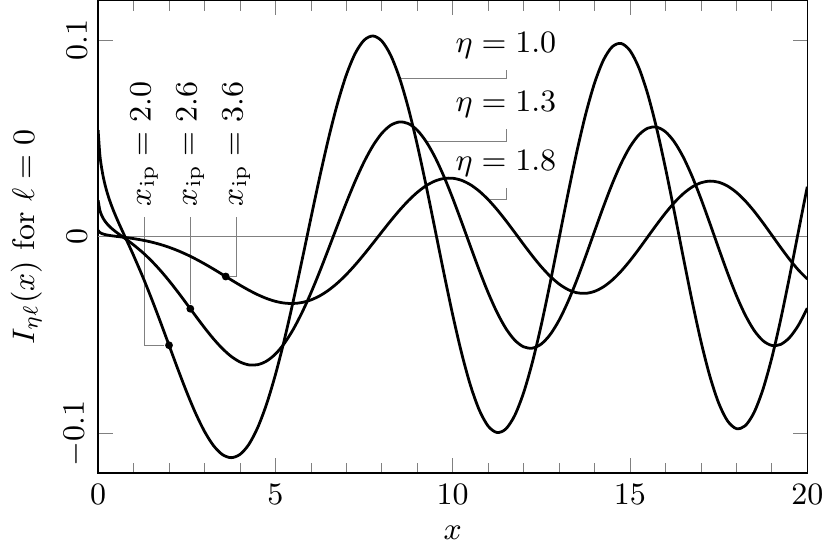}
\caption{\label{fig:plot-coulomb-i}Plots of the regularized Coulomb wave function $I_{\eta\ell}(x)$ for the $s$ wave $(\ell=0)$ and different values of $\eta$.
The inflection points $x_{\rm ip}$ are marked with a black dot. They lie at the same abscissa as for the usual Coulomb functions $F_{\eta\ell}(x)$ and $G_{\eta\ell}(x)$.}
\end{figure}
\par However, the functions $I^\pm_{\eta\ell}(x)$ and $I_{\eta\ell}(x)$ are not asymptotically normalized in the same way as the usual Coulomb functions, as shown in Fig.~\ref{fig:plot-coulomb-i}.
Their wave amplitude is affected by $\eta$ and thus by the energy.
The larger $\eta$, the wider the plateau below the inflection point, as with the regular Coulomb function $F_{\eta\ell}(x)$.
It is interesting that $I_{\eta\ell}(x)$ combines some features of $G_{\eta\ell}(x)$ and $F_{\eta\ell}(x)$ near the origin.
As shown in Fig.~\ref{fig:plot-coulomb-i}, $I_{\eta\ell}(x)$ shows the same singularity as $G_{\eta\ell}(x)$ at $x=0$.
\par The final analytic structure of $G_{\eta\ell}(x)$ can be obtained by averaging the decompositions of $H^+_{\eta\ell}(x)$ and $H^-_{\eta\ell}(x)$ from Eq.~\eqref{eq:CoulombHStructure}.
One gets the analytic decomposition~\cite{Yost1936, Breit1936, Cornille1962, Abramowitz1964, Lambert1969, Humblet1984, Rakityansky2013}  \begin{equation}\label{eq:CoulombGStructure}
\boxed{G_{\eta\ell}(x) = \frac{\E^{2\eta\pi}-1}{\pi} \left[g_\ell(\eta) F_{\eta\ell}(x) + \frac{1}{w_{\eta\ell}} I_{\eta\ell}(x)\right]}  \:,
\end{equation}
where $I_{\eta\ell}(x)$ is the modified Coulomb function given by Eq.~\eqref{eq:CoulombI}.
\par Our definition of $I_{\eta\ell}(x)$ has the advantage of being on a par with $F_{\eta\ell}(x)$ regarding the $k$-dependence.
Indeed, Eqs.~\eqref{eq:CoulombFSeries} and~\eqref{eq:CoulombIpm} show that both $F_{\eta\ell}/C_{\eta\ell}k^{\ell+1}$ and $I_{\eta\ell}/C_{\eta\ell}k^{\ell+1}$ are holomorphic in the $k$-plane and behave as a constant around the zero-energy point.
This result has very important consequences in the framework of effective-range functions, as will be seen below.
Moreover, Eq.~\eqref{eq:CoulombGStructure} provides a clear identification of the singularities of the irregular Coulomb function $G_{\eta\ell}(x)$.

\section{Effective-range functions\label{sec:ERFs}}
The standard ERF is derived from the analysis of the Coulomb wave functions in the complex plane of the energy, when a short-range potential is added to the Coulomb potential.
At the end of the section, the possible alternatives to the standard ERF are presented from the theoretical point of view.

\subsection{Standard effective-range function\label{sec:UsualERF}}
In this section, we show that the analytic decomposition~\eqref{eq:CoulombGStructure} is responsible for the expression of the standard effective-range function.
First, we assume that the potential $V(r)$ describing the interaction between the charged particles is modified by a short-range contribution of nuclear origin.
Therefore, the wave function $u_{k\ell}(r)$ --- that is equal to $F_{\eta\ell}(kr)$ for a pure Coulomb field --- is altered in the nuclear region, whereas it merely acquires the phase shift $\delta_\ell(k)$ in the far-field region.
The phase shift $\delta_\ell(k)$ is defined by the asymptotic behavior of the wave function $u_{k\ell}(r)$ up to a global normalization factor by~\cite{Breit1936, Bethe1949, Landau1944, Blatt1948, Chew1949, Jackson1950, Messiah1961, Gottfried1966, Joachain1979, Newton1982, Haeringen1985, Humblet1990}
\begin{equation}\label{eq:PhaseShiftDef}
u_{k\ell}(r) \xrightarrow{r\rightarrow\infty} F_{\eta\ell}(kr) \cos\delta_\ell(k) + G_{\eta\ell}(kr) \sin\delta_\ell(k)  \:.
\end{equation}
The phase shift follows from the continuity of the logarithmic derivative between the complete wave function $u_{k\ell}(r)$ and Eq.~\eqref{eq:PhaseShiftDef}
\begin{equation}\label{eq:DerivativeContinuity}
\frac{\partial_r u_{k\ell}(R)}{u_{k\ell}(R)} = \frac{\partial_r F_{\eta\ell}(kR) \cos\delta_\ell + \partial_r G_{\eta\ell}(kR) \sin\delta_\ell}{F_{\eta\ell}(kR) \cos\delta_\ell + G_{\eta\ell}(kR) \sin\delta_\ell}  \:.
\end{equation}
The matching point $r=R$ is chosen far enough for the phase shift to converge to an  $R$-independent value.
From Eq.~\eqref{eq:DerivativeContinuity}, one gets the phase shift expressed as a ratio of Wronskians involving the complete wave function $u_{k\ell}(r)$ and the Coulomb functions
\begin{equation}\label{eq:CotgDelta1}
\cot\delta_\ell(k) = \frac{\Wr[G_{\eta\ell}(kr), u_{k\ell}(r)]_R}{\Wr[u_{k\ell}(r), F_{\eta\ell}(kr)]_R}  \:,
\end{equation}
where the notation of the Wronskian determinant is defined as
\begin{equation}
\Wr[f(x),g(x)] = f(x) \,\der{g}{x}(x) - \der{f}{x}(x) \,g(x)  \:.
\end{equation}
Using the analytic decomposition~\eqref{eq:CoulombGStructure} of $G_{\eta\ell}(kr)$ the Wronskian in the numerator of Eq.~\eqref{eq:CotgDelta1} splits into two terms
\begin{equation}\label{eq:CotgDelta2}\begin{split}
\cot\delta_\ell(k) & = \frac{\E^{2\eta\pi}-1}{\pi}   \\
 & \times\left[\frac{1}{w_{\eta\ell}}\frac{\Wr[I_{\eta\ell}(kr), u_{k\ell}(r)]_R}{\Wr[u_{k\ell}(r), F_{\eta\ell}(kr)]_R} - g_\ell(\eta)\right]  \:.
\end{split}\end{equation}
The ratio of Wronskians in Eq.~\eqref{eq:CotgDelta2} involves $I_{\eta\ell}(kr)$ and $F_{\eta\ell}(kr)$, which both behave as $C_{\eta\ell}k^{\ell+1}$ in the neighborhood of $k=0$ as shown in Sec.~\ref{sec:CoulombStructure}.
Therefore, the ratio
\begin{equation}\label{eq:WronkianRatio}
\frac{\Wr[I_{\eta\ell}(kr), u_{k\ell}(r)]_R}{\Wr[u_{k\ell}(r), F_{\eta\ell}(kr)]_R} = w_{\eta\ell}\left[\frac{\pi\cot\delta_\ell(k)}{\E^{2\eta\pi}-1} + g_\ell(\eta)\right]
\end{equation}
is analytic at zero energy and behaves as a constant near $k=0$, due to the cancellation of $C_{\eta\ell}k^{\ell+1}$ from $I_{\eta\ell}(kr)$ and $F_{\eta\ell}(kr)$.
Indeed, all the possible poles of $u_{k\ell}(r)$ in the $k$-plane will simplify in the ratio. 
Consequently, one can define the Coulomb-modified effective-range function based on Eq.~\eqref{eq:WronkianRatio}~\cite{Bethe1949, Landau1944, Blatt1948, Chew1949, Jackson1950, Cornille1962, Lambert1969, Joachain1979, Newton1982, Humblet1990} as
\begin{equation}\label{eq:UsualERF}
\varkappa_\ell(k) = \frac{2w_{\eta\ell}}{\ell!^2\anbohr^{2\ell+1}} \big[\Delta_\ell(k) + g_\ell(\eta)\big]  \:,
\end{equation}
where $\Delta_\ell(k)$ is the reduced effective-range function defined by
\begin{equation}\label{eq:ReducedERF}
\Delta_\ell(k) = \frac{\pi\cot\delta_\ell(k)}{\E^{2\eta\pi}-1}  \:,
\end{equation}
that is discussed in further details in Sec.~\ref{sec:ReducedERF}.
The function $\Delta_\ell(k)$ has been originally defined in Ref.~\cite{Sparenberg2017} with an additional factor $2/\anbohr$, but we omit it in this paper for convenience.
The coefficient in Eq.~\eqref{eq:UsualERF} ensures that it reduces to the effective-range function of the neutral case
\begin{equation}\label{eq:NeutralERF}
\varkappa_\ell(k) = k^{2\ell+1} \cot\delta_\ell(k)  \:,
\end{equation}
for vanishing charges ($\anbohr^{-1}\rightarrow 0$)~\cite{Bethe1949, Landau1944, Lambert1969, Hamilton1973}.
\par The function $\varkappa_\ell(k)$ in Eq.~\eqref{eq:UsualERF} being analytic at $E=0$, it has a useful series expansion in powers of the energy at this point~\cite{Landau1944, Bethe1949, Blatt1948, Chew1949, Jackson1950, Lambert1969, Hamilton1973, Joachain1979, Kok1980, Newton1982, Haeringen1985}, i.e., the effective-range expansion, which is usually written as
\begin{equation}\label{eq:UsualERE}
\varkappa_\ell(k) = -\frac{1}{\alpha_\ell} + \frac{r_\ell}{2}k^2 + \bigo(k^4)  \:,
\end{equation}
where $\alpha_\ell$ is the scattering length and $r_\ell$ is the effective range.
Higher-order terms in Eq.~\eqref{eq:UsualERE} also exist, see~\cite{Naisse1977, Kok1980, Rakityansky2013, Midya2015b}, but they are not discussed in this paper.

\subsection{Properties of Bethe's g function\label{sec:BetheAnalysis}}
Bethe's function $g_\ell(\eta)$ is undoubtedly one of the most important functions of the effective-range theory of charged particles, as evidenced by its presence in Eq.~\eqref{eq:UsualERF}.
\par Indeed, the analyticity of the traditional ERF $\varkappa_\ell(k)$ in Eq.~\eqref{eq:UsualERF} implies that any singular structure in $g_\ell(\eta)$ is reflected on the reduced ERF $\Delta_\ell(k)$~\cite{Sparenberg2017}.
This is why it is so important to begin the study of $\Delta_\ell(k)$ with that of $g_\ell(\eta)$.
\par As a reminder, the function $g_\ell(\eta)$ is defined by Eqs.~\eqref{eq:BetheGpm} and~\eqref{eq:BetheG} as
\begin{equation}\label{eq:FullBetheG}
g_\ell(\eta) = \frac{\psi(\ell+1+\I\eta) - \ln(\I\eta) + \psi(\ell+1-\I\eta) - \ln(-\I\eta)}{2}  \:.
\end{equation}
It is worth noting that Eq.~\eqref{eq:FullBetheG} is not the usual function $g_\ell(\eta)$ that is defined in the literature, especially regarding the $\ell$-dependence.
In Refs.~\cite{Bethe1949, Cornille1962, Lambert1969, Hamilton1973, Humblet1990}, it is often denoted as $h(\eta)$ or $g(\eta)$ and does not depend on $\ell$
\begin{equation}\label{eq:LittBetheG}
g(\eta) = \frac{\psi(1+\I\eta)  + \psi(1-\I\eta)}{2} - \ln\eta  \:.
\end{equation}
As will be seen below, the additional dependence on $\ell$ in Eq.~\eqref{eq:FullBetheG} makes no difference in the ERF from the analytic point of view, except on the effective-range parameters of course.
\par Besides, the function $g_\ell(\eta)$ given by Eq.~\eqref{eq:FullBetheG} reduces to Eq.~\eqref{eq:LittBetheG} for the $s$ wave when $\ell=0$.
Our formulation~\eqref{eq:FullBetheG} has the advantage of being closer to the definition~\eqref{eq:CoulombG} of the function $G_{\eta\ell}(kr)$ from which $g_\ell(\eta)$ originates.
\begin{figure}[ht]
\inputpgf{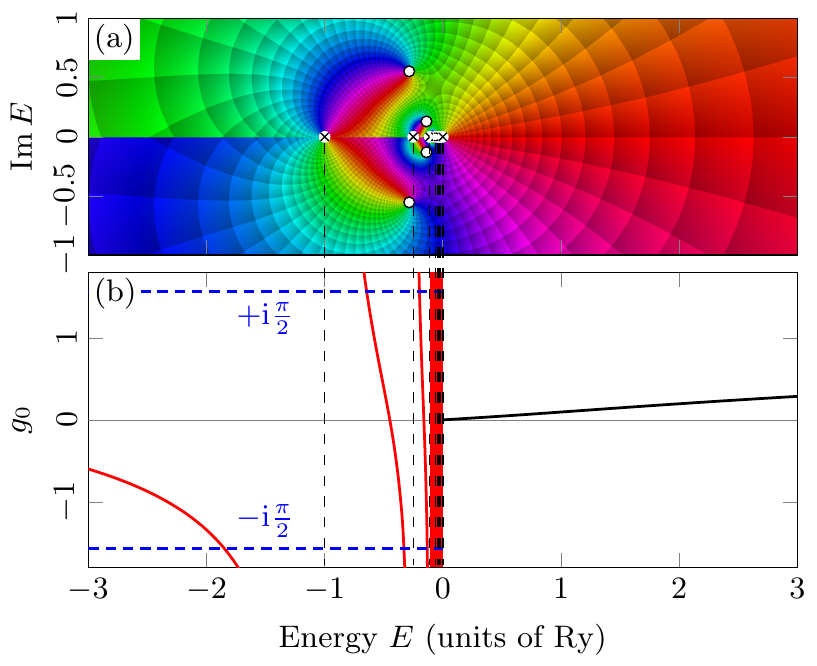}
\caption{\label{fig:plot-g-bethe}(Color online) Representations of the principal branch of $g_\ell(\eta)$ for $\ell=0$ (a) in the complex plane of the energy $E$ and (b) along the real $E$-axis.
In (a), the conventional branch cut lies along the negative real $E$-axis.
Throughout this paper, real functions are depicted in solid black. Where functions are complex, the real part is shown in solid red (dark gray) and the imaginary part in dashed blue.
In (b), the imaginary part is either $+\I\pi/2$ (if $\arg E=+\pi$) or $-\I\pi/2$ (if $\arg E=-\pi$).}
\end{figure}
\par The function $g_\ell(\eta)$ is shown in Fig.~\ref{fig:plot-g-bethe} for $\ell=0$.
It is a real function at positive energy but complex valued everywhere else, especially on the negative real $E$-axis.
\par One of the most noticeable structures is the array of poles at negative energies due to the digamma functions $\psi(\ell+1\pm\I\eta)$.
In the energy plane, they are located at
\begin{equation}\label{eq:CoulombPoles}
k_{{\rm C},n} = \frac{\pm\I}{(n+\ell+1)\anbohr}  \quad\text{and}\quad  E_{{\rm C},n} = \frac{\hbar^2}{2m}k_{{\rm C},n}^2  \:,
\end{equation}
for $n\in\{0, 1, \ldots\}$, so that $E=0$ is an accumulation point of poles and thus an essential singularity of $g_\ell(\eta)$.
Remarkably, the poles of Eq.~\eqref{eq:CoulombPoles} lie at the same energies as the hydrogen-like levels due to the $\ell$-dependence of $g_\ell(\eta)$.
These poles are not related to bound states since the corresponding poles of the reduced ERF $\Delta_\ell(k)$ have no such interpretation.
\par In addition to the poles, the function $g_\ell(\eta)$ also shows infinitely many zeroes near the poles accumulating at $E=0$, as shown in Fig.~\ref{fig:plot-g-bethe}(a).
The pole-zero screening explains why the function becomes suddenly smooth at positive energy.
\par Such a smoothness of $g_\ell(\eta)$ at positive energy may suggest that it is analytic at $E=0$ and can be expanded in series at this point.
If such an expansion was found, the function $g_\ell(\eta)$ could be merely omitted from the Coulomb-modified ERF~\eqref{eq:UsualERF} as proposed in~\cite{Sparenberg2017}, since $\Delta_\ell(k)$ would also be analytic.
However, because of the essential singularity at $E=0$, the function $g_\ell(\eta)$ is not analytic at this point, meaning that no Taylor expansion is expected to converge in a finite neighborhood of $E=0$.
\par On the other hand, according to Refs.~\cite{Olver2010, Abramowitz1964}, the digamma function $\psi(z)$ has an asymptotic Stirling expansion at $\abs{z}\rightarrow\infty$ in all directions except the poles $\abs{\arg z} < \pi-\varepsilon$ ($\varepsilon>0$).
The corresponding asymptotic low-energy behavior of $g_\ell(\eta)$ reads
\begin{equation}\label{eq:LowEnergyBetheG}
g_\ell(\eta) \sim -\sum_{n=1}^\infty \frac{B_{2n}}{2n} (\I\anbohr k)^{2n} + \sum_{s=0}^\ell \frac{(\anbohr k)^2 s}{1 + (\anbohr ks)^2}  \:,
\end{equation}
for $\abs{E}\ll 1~\Ryd$ and $\abs{\arg E} < \pi-\varepsilon$.
The coefficients $B_{2n}$ in Eq.~\eqref{eq:LowEnergyBetheG} are the Bernoulli numbers.
They are known to dramatically increase with $n$~\cite{Olver2010, Abramowitz1964}
\begin{equation}\label{eq:LargeBernoulli}
\frac{B_{2n}}{2n} \sim (-1)^{n+1}\frac{2\,\Gamma(2n)}{(2\pi)^{2n}} \quad\text{for}~n\rightarrow\infty  \:.
\end{equation}
Such an increase reduces to zero the radius of convergence of~\eqref{eq:LowEnergyBetheG}.
\par Regarding the other kinds of rational expansion, the logarithmic branch cut seen in Fig.~\ref{fig:plot-g-bethe}(a) along the negative real $E$-axis will prevent the approximants from converging to $g_\ell(\eta)$.
The logarithmic component in $g_\ell(\eta)$ is also evidenced by its high-energy behavior
\begin{equation}\label{eq:HighEnergyBetheG}
g_\ell(\eta) = \psi(\ell+1) - \ln(\eta) + \bigo(\eta)  \:,
\end{equation}
for $\abs{E}\gg 1~\Ryd$.
The behavior~\eqref{eq:HighEnergyBetheG} also means that $g_\ell(\eta)$ is a flat function at significantly higher energies than the nuclear Rydberg.
\par Besides the function $g_\ell(\eta)$, there are other important functions defined in the literature (see~\cite{Cornille1962, Hamilton1973, Haeringen1985, Kok1980, Humblet1984}): namely the functions $h^\pm_\ell(\eta)$.
As for $g_\ell(\eta)$, they are typically encountered in their $\ell$-independent version.
We define it with a dependence in $\ell$
\begin{equation}\label{eq:HamiltonH}
h^\pm_\ell(\eta) = \psi(\pm\I\eta) + \frac{1}{\pm2\I\eta} - \ln(\pm\I\eta) + \sum_{s=0}^\ell \frac{s}{s^2+\eta^2}  \:,
\end{equation}
so that they are on an equal footing with $g_\ell(\eta)$.
This function is discussed further in Sec.~\ref{sec:HamiltonERF}.
\par The important properties of $h^\pm_\ell(\eta)$ are~\cite{Cornille1962, Hamilton1973}
\begin{equation}\label{eq:HamiltonHtoG}\begin{cases}
\frac{h^+_\ell(\eta) + h^-_\ell(\eta)}{2}   = g_\ell(\eta)                 \:,\\[5pt]
\frac{h^+_\ell(\eta) - h^-_\ell(\eta)}{2\I} = \frac{\pi}{\E^{2\eta\pi}-1}  \:.
\end{cases}\end{equation}
In other words, $g_\ell(\eta)$ can be looked upon as the real part of either $h^+_\ell(\eta)$ or $h^-_\ell(\eta)$ at positive energy, since they are complex conjugated to each other~\cite{Hamming1973}.
In addition, the imaginary part of $h^+_\ell(\eta)$ at $E>0$ is nothing but the Coulomb factor $\pi/(\E^{2\eta\pi}-1)$, as it appears in Eq.~\eqref{eq:CoulombGStructure}.
\par Finally, it is possible to express the far-field behavior of $I_{\eta\ell}(kr)$ substituting the asymptotic formulae of $F_{\eta\ell}(x)$ and $G_{\eta\ell}(x)$ into Eq.~\eqref{eq:CoulombGStructure} and using the properties~\eqref{eq:HamiltonHtoG} to get for positive energies
\begin{equation}\label{eq:FarFieldCoulombI}\begin{split}
I_{\eta\ell}(x) & \xrightarrow{x\rightarrow\infty} -w_{\eta\ell} \abs{h^+_\ell(\eta)} \\
\times & \sin\left(x - \ell\frac{\pi}{2} - \eta\ln(2x) + \sigma_{\eta\ell} - \arg h^+_\ell(\eta)\right)  \:.
\end{split}\end{equation}

\subsection{Effective-range function by Hamilton \textit{et al.}\label{sec:HamiltonERF}}
Before studying further the function $\Delta_\ell(k)$, it is useful to take a look at other functions envisioned as potential alternatives to the traditional ERF $\varkappa_\ell(k)$.
One of the drawbacks of $\Delta_\ell(k)$ is the presence of expected Coulomb poles at negative energy predicted by the analysis of $g_\ell(\eta)$ in Sec.~\ref{sec:BetheAnalysis}.
These poles symmetrically occur in both the physical ($\Im k>0$) and the unphysical sheet ($\Im k<0$), possibly impairing the study of negative energies with $\Delta_\ell(k)$.
\par However, there is a way to partially overcome the issue by removing the poles of $g_\ell(\eta)$ from the physical sheet.
One idea is to apply the reflection formula of the digamma function~\cite{Abramowitz1964, Olver2010} to $\psi(\ell+1-\I\eta)$ in Eq.~\eqref{eq:FullBetheG}, which is responsible for the poles on the positive imaginary $k$-axis.
For this purpose, we use the property
\begin{equation}\label{eq:BethePsiReflection}
\psi(\ell+1-\I\eta) = \psi(\I\eta) - \I\pi - \frac{2\I\pi}{\E^{2\eta\pi}-1} + \sum_{s=1}^\ell \frac{1}{s-\I\eta}  \:,
\end{equation}
where the remaining digamma function $\psi(\I\eta)$ only has poles in the unphysical sheet.
Inserting Eq.~\eqref{eq:BethePsiReflection} into the expression~\eqref{eq:FullBetheG} of $g_\ell(\eta)$ provides the relation
\begin{equation}\label{eq:BetheGtoH}
g_\ell(\eta) = h^+_\ell(\eta) - \frac{\I\pi}{\E^{2\eta\pi}-1}  \:,
\end{equation}
where one notices the appearance of the function $h^+_\ell(\eta)$ defined in Eq.~\eqref{eq:HamiltonH}.
Indeed, the relation~\eqref{eq:BetheGtoH} directly follows from Eq.~\eqref{eq:HamiltonHtoG}.
\par Furthermore, the decomposition~\eqref{eq:BetheGtoH} leads to new formulation of the usual ERF $\varkappa_\ell(k)$ due to Cornille and Martin~\cite{Cornille1962} and Hamilton \textit{et al.}~\cite{Hamilton1973}
\begin{equation}\label{eq:HamiltonUsualERF}
\varkappa_\ell(k) = \frac{2w_{\eta\ell}}{\ell!^2\anbohr^{2\ell+1}} \bigg[\underbrace{\frac{\pi[\cot\delta_\ell(k)-\I]}{\E^{2\eta\pi}-1}}_{\Delta^+_\ell(k)} + h^+_\ell(\eta)\bigg]  \:,
\end{equation}
where $\Delta^+_\ell(k)$ embeds the remaining exponential term in Eq.~\eqref{eq:BetheGtoH}~\cite{Hamilton1973}.
The calculation being symmetric for $h^+_\ell(\eta)$ and $h^-_\ell(\eta)$, one defines the two notations accordingly
\begin{equation}\label{eq:HamiltonERF}
\Delta^\pm_\ell(k) = \Delta_\ell(k) \mp \frac{\I\pi}{\E^{2\eta\pi}-1} = \frac{\pi[\cot\delta_\ell(k)\mp\I]}{\E^{2\eta\pi}-1}  \:.
\end{equation}
The function $\Delta^+_\ell(k)$ is denoted as $F_0^{-1}$ in~\cite{Hamilton1973} up to the factor $2/\anbohr$, and has been applied more recently to the $^2\mathrm{H} + \alpha$ elastic scattering process by Blokhintsev \emph{et al.}~\cite{Blokhintsev2017a, Blokhintsev2017b}.
\begin{figure}[ht]
\inputpgf{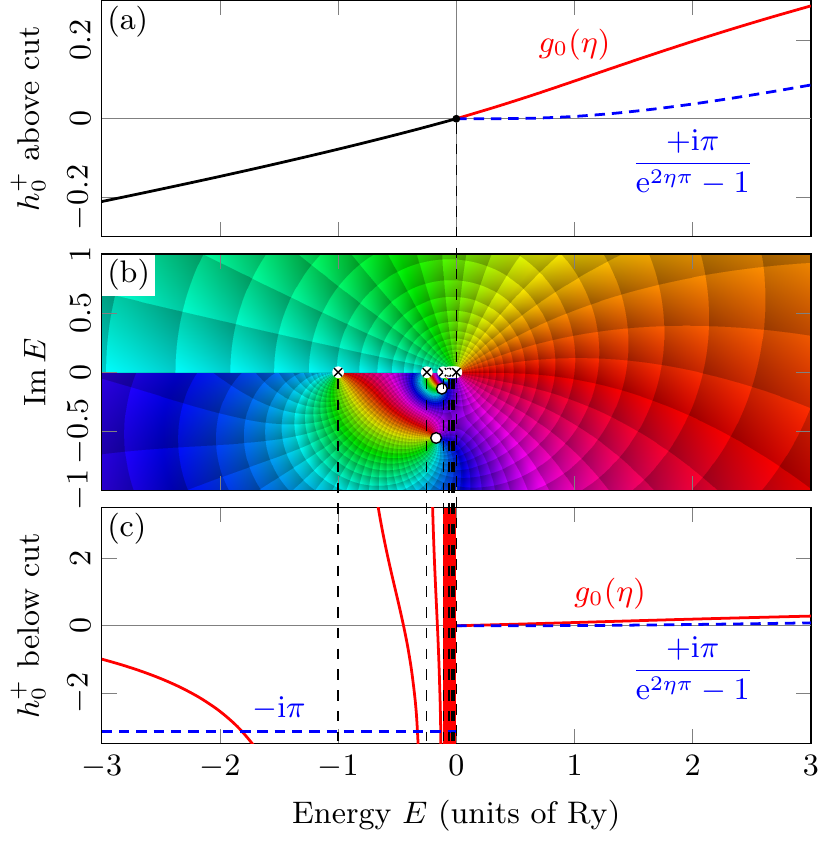}
\caption{\label{fig:plot-h-hamilton}(Color online) Plots of the principal branch of $h^+_\ell(\eta)$ for $\ell=0$ (b) in the complex $E$-plane, (a) along the real $E$-axis in the physical sheet ($\arg E = 0,+\pi$) and (c) along the real axis in the unphysical sheet ($\arg E = 0,-\pi$).
The function is complex valued for $\arg E=0$ but real for $\arg E=+\pi$.
In (c), $h^+_0(\eta)$ is complex valued and shows the Coulomb poles.}
\end{figure}
\par Similarly to $g_\ell(\eta)$ and $\Delta_\ell(k)$, the function $h^+_\ell(\eta)$ is indicative of the behavior of $\Delta^+_\ell(k)$ in the $k$-plane.
As shown in Fig.~\ref{fig:plot-h-hamilton}, $h^+_\ell(\eta)$ --- and thus $\Delta^+_\ell(k)$ --- is complex valued at positive energy, but real on the positive imaginary $k$-axis, that is to say at negative energy above the branch cut ($\arg E=+\pi$).
This property is due to the compensation between the constant imaginary part $-\I\pi/2$ of $\I\pi/(\E^{2\eta\pi}-1)$ and the imaginary part $+\I\pi/2$ of $g_\ell(\eta)$ at $\arg E=+\pi$, as shown in Fig.~\ref{fig:plot-g-bethe}(b).
Below the branch cut ($\arg E=-\pi$), the function $h^+_\ell(\eta)$ is complex valued and has the same Coulomb poles as $g_\ell(\eta)$, but not the same zeroes.
Therefore, although it is smooth near $E=0$ in Fig.~\ref{fig:plot-h-hamilton}(a), the function $\Delta^+_\ell(k)$ is still not analytic at $E=0$. In this respect, the use of $\Delta^+_\ell(k)$ as a potential substitute for the traditional ERF $\varkappa_\ell(k)$ is very debatable.
This topic has never been pursued in the literature until now~\cite{Blokhintsev2017a, Blokhintsev2017b, Sparenberg2017}.
\par It should be noted that $\Delta^+_\ell(k)$ has the great advantage of mimicking the denominator of the Coulomb-modified scattering matrix element~\cite{Gottfried1966, Joachain1979, Newton1982}
\begin{equation}\label{eq:CoulombSMatrix}
S_\ell(k) = \E^{2\I\sigma_{\eta\ell}}\,\frac{\cot\delta_\ell(k) + \I}{\cot\delta_\ell(k) - \I} = \E^{2\I\sigma_{\eta\ell}}\,\frac{\Delta^-_\ell(k)}{\Delta^+_\ell(k)}  \:.
\end{equation}
Therefore, the poles of $S_\ell(k)$ are merely given by the zeroes of $\Delta^+_\ell(k)$~\cite{Blokhintsev2017a}.
\par Using a two-term approximation of the effective-range expansion given by Eq.~\eqref{eq:UsualERE} and the ERF of Eq.~\eqref{eq:HamiltonUsualERF}, the equation of bound and resonant states $\Delta^+_\ell(k)=0$ can also be written as
\begin{equation}\label{eq:CoulombBoundState}
w_{\eta\ell} \,h^+_\ell(\eta) = \frac{\ell!^2\anbohr^{2\ell+1}}{2} \left(-\frac{1}{\alpha_\ell} + \frac{r_\ell}{2}k^2\right)  \:,
\end{equation}
to be solved for the unknowns $k$ and $\eta$ using Eq.~\eqref{eq:SommerfeldEta}. \par When searching for bound states numerically, it is more appropriate to solve Eq.~\eqref{eq:CoulombBoundState} in the $k$-plane than in the $E$-plane.
As shown in Fig.~\ref{fig:plot-h-hamilton}(b), the branch cut of the principal-valued function $h^+_\ell(\eta)$ along the negative $E$-axis will prevent most iterative root-finding methods from converging to the bound state.
The practical advantage of the $k$-representation of $h^+_\ell(\eta)$ is the absence of any singularity in the physical sheet.
\par If, in addition, we consider the $s$ wave at energies low enough that the effective-range term $r_0k^2/2$ is negligible, the equation~\eqref{eq:CoulombBoundState} for bound and resonant states takes the form
\begin{equation}\label{eq:CoulombSBoundState}
h^+_0(\eta) = \psi(\I\eta) + \frac{1}{2\I\eta} - \ln(\I\eta) = \frac{-\anbohr}{2\alpha_0}   \:.
\end{equation}
This equation is encountered with Dirac delta-plus-Coulomb potentials, since the effective-range $r_0$ is zero as well as any higher order coefficient~\cite{Albeverio1988}.  \par The solutions of Eq.~\eqref{eq:CoulombSBoundState} can be found graphically in Fig.~\ref{fig:plot-h-hamilton}(a) based on the scattering length $\alpha_0$.
When $\alpha_0$ is positive then the solution can be interpreted as a bound state because $h^+_0(\eta)$ can match $-\anbohr/2\alpha_0$ at negative energy.
Otherwise, if $\alpha_0<0$, the solution is interpreted as a resonance.
In the latter case, the zero of $\Delta^+_\ell(k)$ deviates from the real $E$-axis towards the unphysical sheet so as to follow the level curve defined by $\Im h^+_\ell(\eta) = 0$ in Fig.~\ref{fig:plot-h-hamilton}(b).
This curve is referred to as ``universal'' in Ref.~\cite{Kok1980}, although it is only valid at zeroth-order approximation of the effective-range theory, as in Eq.~\eqref{eq:CoulombSBoundState}.

\subsection{Reduced effective-range function\label{sec:ReducedERF}}
We are now focusing on the properties of the reduced ERF $\Delta_\ell(k)$ and its potential interest in low-energy scattering.
As recently highlighted in~\cite{Sparenberg2017}, a major drawback of the usual ERF $\varkappa_\ell(k)$ is the overwhelming dominance of $g_\ell(\eta)$ upon the phase-shift-dependent part $\Delta_\ell(k)$, which occurs especially with heavy and moderately heavy nuclei.
This imbalance comes from the smallness of the exponential prefactor $\pi/(\E^{2\eta\pi}-1)$ at typical energies encountered in low-energy nuclear scattering experiments ($<10~\Ryd$)
\begin{equation}
\frac{\pi}{\E^{2\eta\pi}-1} \ll g_0(\eta)  \quad\text{for}~\eta\gtrsim 1  \:.
\end{equation}
For instance, at $E=0.1~\Ryd$, the Sommerfeld parameter is $\eta = \sqrt{10}$ and the factor $\pi/(\E^{2\eta\pi}-1)$ is about $10^6$ times smaller than the function $g_0(\eta)$.
The greatness of $g_\ell(\eta)$ is a potential problem while interpolating $\varkappa_\ell(k)$ because it may conceal the structures in $\Delta_\ell(k)$ due to the phase shift.
Therefore, the addition of $g_\ell(\eta)$ could lead to an underfitting of the phase-shift-dependent part $\Delta_\ell(k)$ of the usual ERF, as done in Refs.~\cite{Orlov2016b, Orlov2017b}.
\par One easy way to avoid this drawback is to directly approach the experiment-based function $\Delta_\ell(k)$ by an expansion of the form
\begin{equation}\label{eq:DeltaExpansion}
\Delta_\ell(k) = -g_\ell(\eta) + \frac{\ell!^2\anbohr^{2\ell+1}}{2w_{\eta\ell}}\left[\frac{-1}{\alpha_\ell} + \frac{r_\ell}{2}k^2 + \bigo(k^4)\right]  \:.
\end{equation} It has the advantage of being perfectly consistent with the usual effective-range method~\eqref{eq:UsualERF}.
But now, there is no more risk of superposition between large and small quantities.
\par The peculiar properties of the functions $h^\pm_\ell(\eta)$ and $g_\ell(\eta)$ allow us to go a little further.
As shown in Eq.~\eqref{eq:BetheGtoH}, the only difference between $h^+_\ell(\eta)$ and $g_\ell(\eta)$ is the exponential term behaving like
\begin{equation}\label{eq:ExpSingularity}
\frac{\I\pi}{\E^{2\eta\pi}-1} \sim \I\pi\E^{-2\pi/\anbohr k} \quad\text{as}~E\xrightarrow{>}0  \:.
\end{equation}
Accordingly, the asymptotic expansion of $\I\pi/(\E^{2\eta\pi}-1)$ is zero at $E=0$ (for $E>0$) due to the essential singularity at this point.
Therefore, the function $h^+_\ell(\eta)$ has the same asymptotic expansion~\eqref{eq:LowEnergyBetheG} as $g_\ell(\eta)$
\begin{equation}\label{eq:LowEnergyHamiltonH}
h^+_\ell(\eta) \sim -\sum_{n=1}^{n_{\max}} \frac{B_{2n}}{2n} (\I\anbohr k)^{2n} + \sum_{s=0}^\ell \frac{(\anbohr k)^2 s}{1 + (\anbohr ks)^2}  \:,
\end{equation}
for $E\rightarrow 0$ but in the physical sheet ($\Im k\geq 0$) as the order $n_{\max}$ tends to infinity.
This shows that $h^+_\ell(\eta)$ and $g_\ell(\eta)$ come together smoothly at the origin in the physical sheet, as evidenced by Fig.~\ref{fig:plot-h-hamilton}(a).
Since the two functions $h^+_\ell(\eta)$ and $g_\ell(\eta)$ are similar to $\Delta^+_\ell(k)$ and $\Delta_\ell(k)$ respectively, the above statement also means that $\Delta^+_\ell(k)$ and $\Delta_\ell(k)$ smoothly join at $E=0$ in the physical sheet.
\par To some extent, this property can be exploited to extrapolate low-energy data to negative energies using $\Delta_\ell(k)$ instead of the traditional ERF, as done in Refs.~\cite{Sparenberg2017,Blokhintsev2017a,Blokhintsev2017b}.
Indeed, it is possible that the direct interpolation $\Delta_\ell^\fit(k)$ of the experiment-based function $\Delta_\ell(k)$ locally provides a reasonable estimate of $\Delta^+_\ell(k)$ at negative energies in the physical sheet
\begin{equation}\label{eq:DeltaFitting}\Delta_\ell^\fit(k) \sim \begin{cases}
\Delta_\ell(k)    & \textrm{for}~\arg E = 0     \:,\\[5pt]
\Delta^+_\ell(k)  & \textrm{for}~\arg E = +\pi  \:.
\end{cases}\end{equation}
Therefore, the negative-energy zeroes of $\Delta_\ell^\fit(k)$ can be interpreted as bound states, as long as they are located in a region of low energy (typically $\abs{E}\ll 1~\Ryd$).
\par However, it should be noted that such a method is not guaranteed to provide reliable results, because $\Delta_\ell(k)$ and $\Delta^+_\ell(k)$ are not analytic at $E=0$.  The smoothness of $\Delta^+_\ell(k)$ at $E=0$ in the physical sheet is not enough to consider it as analytic, because of the essential singularity at $E=0$ [see the accumulation of poles in Fig.~\ref{fig:plot-h-hamilton}(c)].
Attempting to interpolate $\Delta_\ell(k)$ by a meromorphic function like a Padé approximant is likely to lead to undetermined behaviors, without possible convergence to $\Delta^+_\ell(k)$.  \par In addition, the analytic continuation of the function $\Delta_\ell(k)$ is multi-valued due to its logarithmic component discussed in Sec.~\ref{sec:BetheAnalysis}.
Such a feature cannot be interpolated by a Padé approximant.
If, though, it is done, the fitted Padé approximant would attempt to accumulate spurious poles on the negative $E$-axis to come closer to the branch cut.
This will be discussed further in Sec.~\ref{sec:Application}.
\par On the other hand, it is possible to quantify the accuracy of the asymptotic expansion~\eqref{eq:LowEnergyHamiltonH}.
This should help to estimate the minimum error made when approaching $\Delta_\ell(k)$ with a polynomial in $E$.
Indeed, despite the convergence of $\Delta_\ell^\fit(k)$ is not expected as the order increases, it may provide useful interpolation in a low enough energy range.
\begin{figure}[ht]\inputpgf{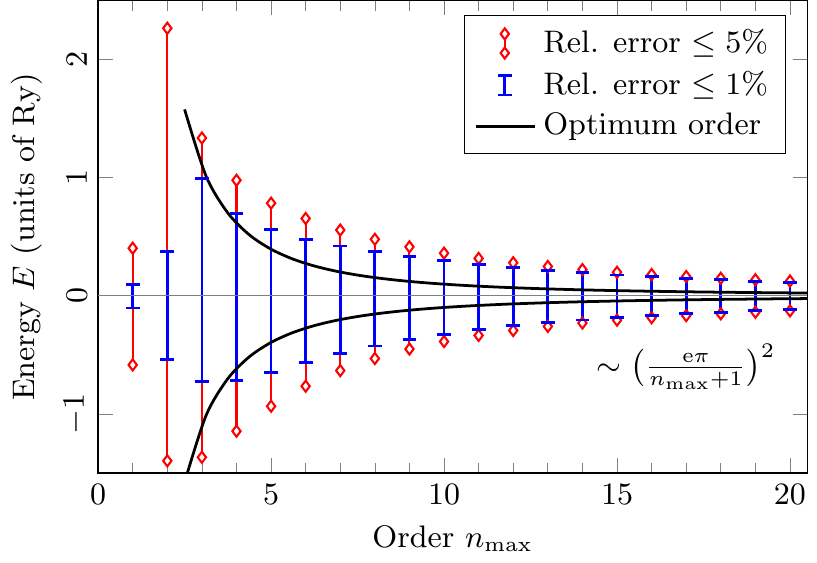}
\caption{\label{fig:approx-h-energy-domain}Energy intervals where the relative error of the truncated Stirling expansion~\eqref{eq:LowEnergyHamiltonH} of $h^+_\ell(\eta)$ (for $\ell=0$) does not exceed $5\%$ and $1\%$.
The expected optimum order $n_{\rm opt}$ of Eq.~\eqref{eq:OptimumStirlingOrder} is shown in solid black.}
\end{figure}
For this purpose, we compute the energy intervals where the relative error in the truncated Stirling series~\eqref{eq:LowEnergyHamiltonH} does not exceed $1\%$ to $5\%$ as a function of the order $n_{\max}$.
The result is shown in Fig.~\ref{fig:approx-h-energy-domain}.
\par The expansion obviously diverges since the region of validity, represented by vertical bars, continues to decrease.
At high orders, an approximate calculation involving Eq.~\eqref{eq:LargeBernoulli} shows that the energy interval roughly decreases like $[\E\pi/(n_{\max}+1)]^2~\Ryd$ as $n_{\max}\rightarrow\infty$, independently of the bound on the relative error.
\par Beyond about $1~\Ryd$, the error is larger than $5\%$ whatever the order $n_{\max}$ in Eq.~\eqref{eq:LowEnergyHamiltonH}.
This suggests that, as long as no data point is known in the interval $[-1,1]~\Ryd$, the interpolation of $\Delta_\ell(k)$ will not be of practical interest at $E<0$. \par However, it is possible to get relatively accurate results if data points are known at energies below $1~\Ryd$.
This case is typically encountered for heavy and moderately heavy particles.
In addition, the intervals in Fig.~\ref{fig:approx-h-energy-domain} allow us to roughly estimate the maximum order $n_{\max}$ before a polynomial interpolation of $\Delta_\ell(k)$ will stray too much from $\Delta^+_\ell(k)$ depending on the energy range considered.
\par Moreover, there is an optimum order that minimizes the error of the asymptotic series in Eq.~\eqref{eq:LowEnergyHamiltonH}.
It also corresponds to the smallest term of this series.
To get it for the $s$ wave, one cancels the logarithmic derivative of the $n$th term in Eq.~\eqref{eq:LowEnergyHamiltonH} using the asymptotic behavior~\eqref{eq:LargeBernoulli}
\begin{equation}\label{eq:OptimumOrderCondition}
\der{}{n}\big[\ln\Gamma(2n) - 2n\ln\abs{2\eta\pi}\big] = 0  \:.
\end{equation}
This is a suitable approximation provided that the sought index $n$ is larger than $1$.
The approximate solution of Eq.~\eqref{eq:OptimumOrderCondition}, rounded to the closest integer, is
\begin{equation}\label{eq:OptimumStirlingOrder}
n_{\rm opt} \simeq \pi\abs{\eta} = \pi\sqrt{\frac{\Ryd}{\abs{E}}}  \:.
\end{equation}
The curve of the optimum order $n_{\rm opt}$ is shown in Fig.~\ref{fig:approx-h-energy-domain}.
Above $n_{\rm opt}$, the Stirling series in Eq.~\eqref{eq:LowEnergyHamiltonH} starts diverging.

\section{Application to proton-proton collision\label{sec:Application}}
In this section, we propose to apply the effective-range theory to the $^1S_0$ elastic scattering of two protons.
Indeed, this two-body system is of historical importance and is greatly documented in the literature, especially in Refs.~\cite{Breit1936, Landau1944, Blatt1948, Bethe1949, Chew1949, Jackson1950, Naisse1977, Kok1980, Bergervoet1988, Mukha2010}.
This section is divided into two parts: the first one is about the graphical representation of the previously discussed effective-range functions $\Delta_\ell(k)$, $\Delta^+_\ell(k)$ and $\varkappa_\ell(k)$ at real and complex energies, and the second one is about the practical use of the reduced ERF $\Delta_\ell(k)$~\cite{Sparenberg2017} in the framework of proton-proton collision.
As a reminder, the orders of magnitude for the proton-proton system are mainly governed by the nuclear Rydberg energy: $1~\Ryd = 12.49~\mathrm{keV}$~\cite{Bethe1949}.

\subsection{Effective-range functions in the \texorpdfstring{$\boldsymbol{E}$}{E}-plane\label{sec:PPFunctions}}
In order to reproduce the phase shift $\delta_\ell(k)$ and the related quantities for the proton-proton scattering, we resort to the square-well model.
We assume the total potential $V(r)$ to be constant in the short-range region $r\leq R$
\begin{equation}V(r) = \begin{cases}
V_0                      & \text{if}~r\leq R  \:,\\[3pt]
\frac{\alpha\hbar c}{r}  & \text{if}~r> R     \:,
\end{cases}\end{equation}
with typically negative $V_0$.
This simple model should be sufficient to describe the functions of interest at relatively low energy, i.e., below about $5~\mathrm{MeV}$ for proton-proton.
\par Such an approach is similar to what has recently been done by Blokhintsev \textit{et al.} in~\cite{Blokhintsev2017a}. However, we assume the additional potential compensates for Coulomb interaction in the nuclear region $r\leq R$.
This provides a total potential $V(r)$ that is both simple and practical.
This choice has no consequences at low energy, but it modifies the high-energy limit of the effective-range functions.
\par The square-well model has the advantage of being exactly solvable.
This will be quite useful in the following to perform the analytic continuation of the functions to the complex $E$-plane.
In the short-range region, the wave function $u_{k\ell}(r)$ is described by the spherical Bessel functions $j_\ell(z)$~\cite{Messiah1961, Gottfried1966, Joachain1979, Newton1982}
\begin{equation}
u_{k\ell}(r) = qr\,j_\ell(qr)  \quad\text{for}~r\leq R  \:,
\end{equation}
where $q$ is the local wave number given by
\begin{equation}
q = \sqrt{k^2 - \frac{2m}{\hbar^2}V_0}  \:.
\end{equation} \par Then, we compute $\cot\delta_\ell(k)$ from Eq.~\eqref{eq:CotgDelta1} using the recurrence properties of the Coulomb wave functions to compute their derivatives~\cite{Olver2010}.
The computation is done in the \textsc{Wolfram Mathematica} software~\cite{Wolfram1999} using the implementation of the confluent hypergeometric functions $M(a,b,z)$ and $U(a,b,z)$ defined for complex arguments. \par Finally, the parameters $R$ and $V_0$ of the potential well are fitted to reproduce the effective-range parameters $\alpha_0$ and $r_0$ for the proton-proton $^1S_0$ channel as given by Refs.~\cite{Naisse1977, Bergervoet1988}
\begin{equation}\label{eq:PPEffRangeParams}\begin{cases}
\alpha_0  = -7.81~\mathrm{fm}  \:,\\
r_0       = 2.79~\mathrm{fm}   \:.
\end{cases}\end{equation}
We obtain the parameters
\begin{equation}\label{eq:PPSquareWellParams}\begin{cases}
R   = 2.8~\mathrm{fm}       \:,\\
V_0 = -10.66~\mathrm{MeV}   \:.
\end{cases}\end{equation}
In the following subsections, all the phase-shift-related quantities have been computed for the square-well model with the parameters of Eq.~\eqref{eq:PPSquareWellParams}.

\subsubsection{Proton-proton reduced effective-range function\label{sec:PPReducedERF}}
\begin{figure}[ht]
\inputpgf{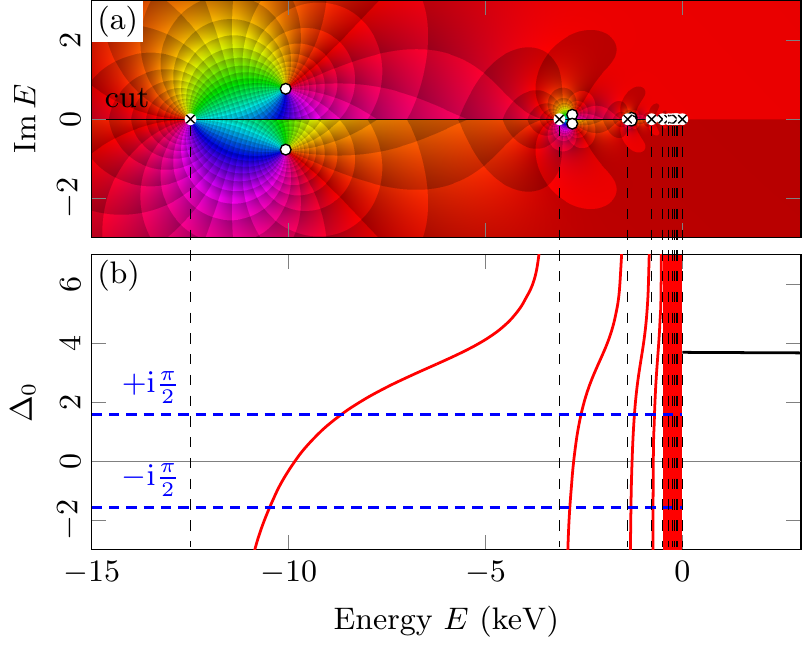}
\caption{\label{fig:plot-pp-delta-fct-low}(Color online) Function $\Delta_0(E)$ for the proton-proton $^1S_0$ scattering (a) in the complex $E$-plane and (b) on the real $E$-axis.
The function is real and positive at $E>0$ but complex-valued at $E<0$. In (a), the branch cut along the negative $E$-axis is depicted by a black line.
The imaginary part around the cut is either $-\I\pi/2$ (if $\arg E=+\pi$) or $+\I\pi/2$ (if $\arg E=-\pi$).}
\end{figure}

\begin{figure}[ht]
\inputpgf{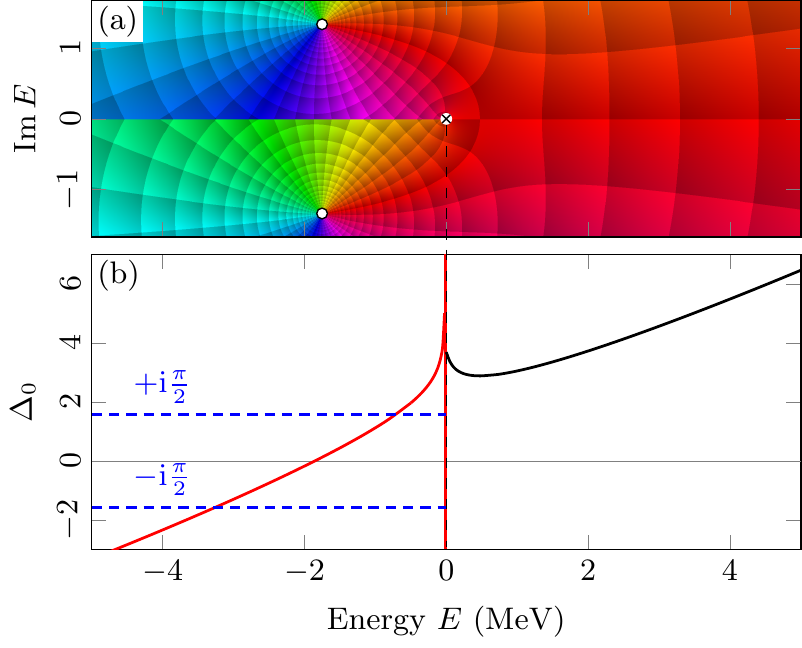}
\caption{\label{fig:plot-pp-delta-fct-high}(Color online) Same as Fig.~\ref{fig:plot-pp-delta-fct-low} but at higher energies.
At this scale, the Coulomb poles look merged at $E=0$. The two zeroes are located at about $E = (-1.75\pm 1.33\,\I)~\mathrm{MeV}$.}
\end{figure}

The reduced ERF $\Delta_0(E)$ of the proton-proton $s$ wave is represented in the complex plane of the energy in Figs.~\ref{fig:plot-pp-delta-fct-low} and~\ref{fig:plot-pp-delta-fct-high}.

\par The function $\Delta_0(E)$ is shown at two different scales because of the large disparity of the orders of magnitude for this system.
Indeed, $\Delta_0(E)$ is almost linear beyond roughly $2~\mathrm{MeV}$ in Fig.~\ref{fig:plot-pp-delta-fct-high}(b), but also has poles below $1~\Ryd = 12.49~\mathrm{keV}$ accumulating at $E=0$ on the negative real $E$-axis in Fig.~\ref{fig:plot-pp-delta-fct-low}.
\par It should be noted that these poles exactly correspond to the Coulomb poles of $g_0(\eta)$ given by Eq.~\eqref{eq:CoulombPoles}. The Coulombic nature of the poles of $\Delta_0(E)$ can be checked by observing that they are independent of the nuclear potential depth $V_0$.
Besides, the function $\Delta_0(E)$ also shows an accumulation of zeroes at $E=0$ that are compensating for the poles, hence the smooth behavior at $E>0$.
The two high-energy zeroes of $\Delta_0(E)$ in Fig.~\ref{fig:plot-pp-delta-fct-high}(a) are located at about $E = (-1.75\pm 1.33\,\I)~\mathrm{MeV}$, and correspond to points where the scattering matrix from Eq.~\eqref{eq:CoulombSMatrix} amounts to $-1$ up to a pure Coulomb phase.
Unlike the Coulomb poles, all these zeroes depend on the parameters of the nuclear potential.  \par In addition, $\Delta_0(E)$ possesses a branch cut clearly visible in Fig.~\ref{fig:plot-pp-delta-fct-high}(a) and highlighted by a black line in Fig.~\ref{fig:plot-pp-delta-fct-low}(a).
The cut stops at $E=0$ and the function is smooth at positive energy. \par Furthermore, the function $\Delta_0(E)$ shows two very different structures for positive and negative real energies in Fig.~\ref{fig:plot-pp-delta-fct-low}(b) due to the essential singularity at this point.
This could suggest that $\Delta_0(E)$ is piecewise defined. However, Fig.~\ref{fig:plot-pp-delta-fct-low}(a) shows that there is no such thing, because these two pieces belong to the same Riemann surface through analytic continuation.
\par Finally, all these low-energy structures confirm that $\Delta_0(E)$ behaves as predicted by the expansion~\eqref{eq:DeltaExpansion} of the effective-range theory.

\subsubsection{Proton-proton effective-range function by Hamilton \textit{et al.}\label{sec:PPHamiltonERF}}
\begin{figure}[ht]
\inputpgf{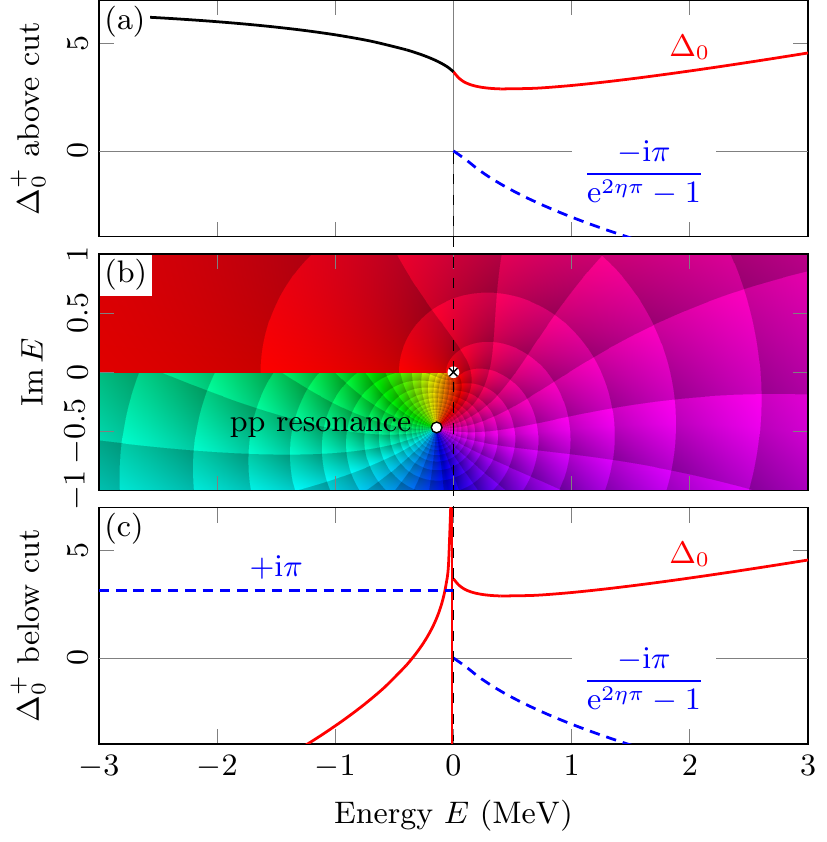}
\caption{\label{fig:plot-pp-deltaplus-high}(Color online) Proton-proton function $\Delta^+_0(E)$ (b) in the complex $E$-plane, (a) on the real $E$-axis in the physical sheet and (c) in the unphysical sheet.
The function is real for $\arg E=+\pi$. Its real part reduces to $\Delta_0(E)$ at $E>0$.
In (c), the Coulomb poles at $\arg E=-\pi$ are indistinguishable at this scale.}
\end{figure}

The ERF $\Delta^+_0(E)$ by Hamilton \textit{et al.}~\cite{Hamilton1973} for the two-proton system is shown in Fig.~\ref{fig:plot-pp-deltaplus-high} at relatively high energies with respect to the nuclear Rydberg energy of $12.49~\mathrm{keV}$.
The plot of $\Delta^+_0(E)$ at low energies ($\abs{E}<1~\Ryd$) is very similar to Fig.~\ref{fig:plot-pp-delta-fct-low}(a), but without poles or zeroes in the physical sheet ($\Im k>0$).
Of course, $\Delta^+_0(E)$ has Coulomb poles and zeroes in the unphysical sheet, as well as the function $h^+_0(\eta)$.
\par As discussed in Sec.~\ref{sec:HamiltonERF}, the function $\Delta^+_0(E)$ features zeroes corresponding to poles of the $S_0$ matrix element.
Indeed, one notices a zero in Fig.~\ref{fig:plot-pp-deltaplus-high}(b) located at
\begin{equation}\label{eq:PPResonance}
E_{\rm res,pp} \simeq (-142 - 467\,\I)~\mathrm{keV}  \:,
\end{equation}
that coincides with the proton-proton broad resonance referred to in literature~\cite{Kok1980, Mukha2010}.
Other nontrivial zeroes exist in the unphysical sheet at low negative energies ($\abs{E}<1~\Ryd$), but, being very close to the negative $E$-axis, they are interpreted in Ref.~\cite{Kok1980} as antibound states.
\par Furthermore, $\Delta^+_0(E)$ is smooth in the physical sheet near $E=0$, although it might be unclear in Fig.~\ref{fig:plot-pp-deltaplus-high}(a).
In fact, the function has a inflection point at about $E=1.1~\Ryd = 13.9~\mathrm{keV}$ very close to the origin at this scale, hence the impression of an angular point at $E=0$.
\par The potential interest of $\Delta^+_0(E)$ to extrapolate experimental data to negative energies in the physical sheet is obvious in Fig.~\ref{fig:plot-pp-deltaplus-high}(a).
Even though the essential singularity at $E=0$ prevents the interpolation functions from converging everywhere in the $E$-plane.

\subsubsection{Proton-proton usual effective-range function\label{sec:PPUsualERF}}
\begin{figure}[ht]
\inputpgf{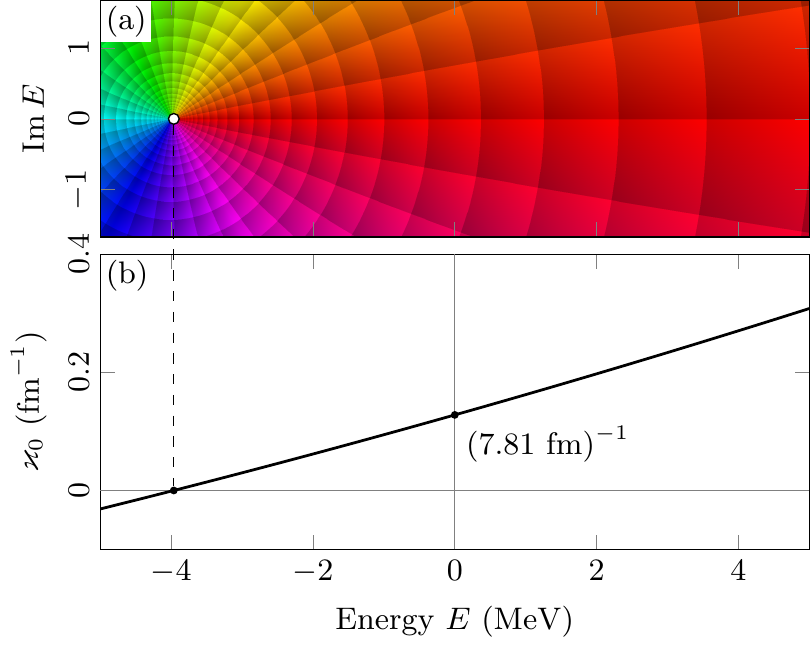}
\caption{\label{fig:plot-pp-erf-high}(Color online) Standard ERF $\varkappa_0(E)$ for the proton-proton $^1S_0$ collision (a) in the complex $E$-plane and (b) on the real $E$-axis.
The zero is located at about $E\simeq-3.8~\mathrm{MeV}$.} \end{figure}

The last function to show for the proton-proton $^1S_0$ wave is the traditional ERF $\varkappa_0(E)$.
It is plotted in the complex $E$-plane in Fig.~\ref{fig:plot-pp-erf-high}.
\par The function $\varkappa_0(E)$ is real valued at both positive and negative energy, and it is analytic at $E=0$.
Remarkably, in the two-proton scattering, $\varkappa_0(E)$ is a nearly straight line because the experimentally observed $\bigo(E^2)$ term is very small~\cite{Landau1944, Bethe1949, Blatt1948, Chew1949, Jackson1950, Kok1980}.
\par The negative-energy zero of $\varkappa_0(E)$ shown in Fig.~\ref{fig:plot-pp-erf-high} is located at about
\begin{equation}
E = \frac{\hbar^2}{2m}\frac{2}{\alpha_0r_0} \simeq -3.8~\mathrm{MeV}  \:,
\end{equation}
using the linear approximation of the usual ERF, and the parameters from Eq.~\eqref{eq:PPEffRangeParams}.
\par All these features justify the method based on the traditional ERF expansion, especially for light particles such as protons.
Indeed, the low-energy behavior of the proton-proton $^1S_0$ phase shift is accurately given over a few MeVs by merely two parameters $\alpha_0$ and $r_0$.

\subsection{Use of the reduced effective-range function\label{sec:UseofReducedERF}}
In this section, we consider using the reduced ERF $\Delta_0(E)$ directly to obtain information on negative energies, as proposed in Ref.~\cite{Sparenberg2017}.
The proton-proton $^1S_0$ scattering is still used as a practical example.
\par The following results can still be compared to other scattering systems provided that we refer to the nuclear Rydberg energy.
Indeed, this quantity governs most of the orders of magnitude of energy in the charged-particle scattering.
This is why energies are expressed in $\Ryd$ hereafter.

\subsubsection{Extraction of resonances}
First, we focus on the determination of the proton-proton resonance from experimental data using the function $\Delta_0(E)$.
For this purpose, we interpolate directly $\Delta_0(E)$ with Padé approximants of different orders, as done in Ref.~\cite{Sparenberg2017}.
\par To compute the fitting, 120 data points sampled logarithmically are taken from the square-well model in the range $[10,10^3]~\Ryd \simeq [0.125, 12.5]~\mathrm{MeV}$.
This range has been chosen because it is the closest to the experimental framework of proton-proton collision.
The resulting Padé approximants of orders $[2/1]$, $[4/3]$ and $[6/5]$ are shown in Fig.~\ref{fig:fit-pp-delta-fct}(a).
Our choice of the orders $[(n+1)/n]$ builds on the Refs.~\cite{Hartt1981, Midya2015b}, but different orders do not affect our results.
\par To get the resonance, one has to find the root of the equation $\cot\delta^\fit_0(E) = \I$, that is to say in terms of $\Delta_0(E)$
\begin{equation}\label{eq:DeltaResonance}
\Delta^\fit_0(E) = \frac{\I\pi}{\E^{2\eta\pi}-1}  \:.
\end{equation} Solving Eq.~\eqref{eq:DeltaResonance} numerically with the Padé approximant $[6/5]$ provides a broad resonance at
\begin{equation}\label{eq:PPResonanceDelta}
E_{\rm res,pp} \simeq (-141 - 467\,\I)~\mathrm{keV}  \:,
\end{equation}
very close to~\eqref{eq:PPResonance} that we have found in the square-well model, and previously reported in Refs.~\cite{Mukha2010, Kok1980}.
Such an agreement can be explained by the remoteness of the resonance pole from the nuclear Rydberg energy ($1~\Ryd\simeq 12.5~\mathrm{keV}$) below which the essential singularity of $\Delta_0(E)$ hinders the convergence of the interpolations, as shown in Fig.~\ref{fig:fit-pp-delta-fct}(a).
In addition, being of modulus $\abs{E_{\rm pp,res}} = 0.488~\mathrm{MeV}$, the pole lies in a ring centered at $E=0$ that covers the data points in $[0.125, 12.5]~\mathrm{MeV}$.
Therefore, the fitting successfully provides the resonance pole, although no convergence of the Padé approximant is observed because of the logarithmic branch cut of $\Delta_0(E)$.
Heuristically, the region of the complex $E$-plane where the fitting is reliable turns out to be the sector $\abs{\arg E}\lesssim 3\pi/4$ and $\abs{E}\gtrsim 1~\Ryd$.
Closer to the negative $E$-axis, spurious poles makes the fitting unusable.
\par Finally, we conclude that it is possible to extract narrow or broad resonances from the reduced ERF $\Delta_0(E)$, provided that the corresponding sought poles are not too close to the negative $E$-axis with respect to the nuclear Rydberg energy.

\subsubsection{Extrapolation to negative energies}
\begin{figure}[ht]\inputpgf{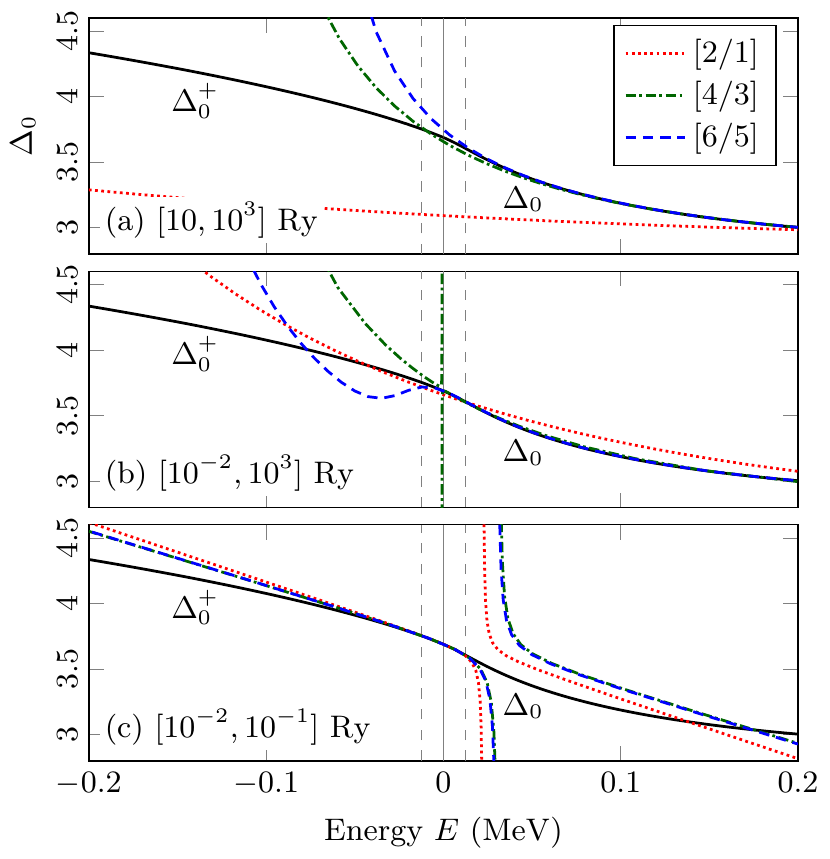}
\caption{\label{fig:fit-pp-delta-fct}Interpolations of $\Delta_0(E)$ at $E>0$ for the proton-proton $^1S_0$ wave with Padé approximants of orders $[2/1]$, $[4/3]$ and $[6/5]$. $\Delta_0(E)$ is sampled logarithmically in 120 points (not shown) on different ranges (a) $[10,10^3]~\Ryd$, (b) $[10^{-2},10^3]~\Ryd$, and (c) $[10^{-2},10^{-1}]~\Ryd$.
The Coulomb range $[-1,1]~\Ryd$ is surrounded by vertical dashed lines. The real part of $\Delta^+_0(E)$ (for $\arg E=+\pi$) is shown in solid black.}
\end{figure}
Now, we consider using the function $\Delta_0(E)$ to extract information on the bound states.
Although the two-proton system has no bound state, it is useful to study how the interpolation of $\Delta_0(E)$ may extend to negative energies, especially, in which circumstances it approaches the function $\Delta^+_0(E)$, as predicted in Eq.~\eqref{eq:DeltaFitting}.
\par As previously, we take 120 sample points of $\Delta_0(E)$ computed in the square-well model to fit the Padé approximants.
Three sampling intervals are envisioned to reproduce different experimental situations.  Case (a) is the typical situation encountered for proton-proton collision: no experimental data is known below about $10~\Ryd\simeq 0.125~\mathrm{MeV}$.
Case (c) is more likely encountered with heavy or moderately heavy nuclei, for which the Rydberg energy is relatively high compared to nuclear energies.
Case (b) is intermediate because involving both low and high energy data with respect to the Rydberg energy.
\par In case (a), 120 data points are sampled in the interval $[10, 1000]~\Ryd \simeq [0.125, 12.5]~\mathrm{MeV}$.
One notices in Fig.~\ref{fig:fit-pp-delta-fct}(a) that none of the Padé approximants is reliable in the Coulomb range, i.e., below $1~\Ryd$.
The high-order Padé approximants $[4/3]$ and $[6/5]$ diverge near the inflection point located at about $1.1~\Ryd$.
Beyond that point, spurious poles appear at negative energy (not visible in Fig.~\ref{fig:fit-pp-delta-fct}).
This means that the $\Delta_0$-based extrapolation is not indicated for light particle systems such as protons.
\par In case (b), we consider a fictitious situation with 120 data points in the interval $[0.01, 1000]~\Ryd$.
Although the Padé approximants are closer to $\Delta^+_0(E)$, they still diverge at negative energy.
In addition, the Padé approximant $[4/3]$ shows a spurious pole very close to $E=0$ in the negative-energy Coulomb range.
Such spurious poles are likely due to the negative-energy branch cut of $\Delta_0(E)$ which prevents the approximants from converging.
From this point of view, additional data are not helpful and create more constraints that the Padé approximants cannot follow anyway.
\par Finally, in case (c), 120 data points are sampled in $[0.01, 0.1]~\Ryd \simeq [0.125, 1.25]~\mathrm{keV}$.
Interestingly, the fitted Padé approximants are significantly closer to $\Delta^+_0(E)$ up to about $-2~\Ryd$.
\par Spurious poles at positive energy in case (c) make the Padé approximants unusable for $E\gtrsim 1~\Ryd$.
This seems to show that, in general, one has to choose between fitting at higher or lower energies than about $1~\Ryd$.
\par The Padé approximants in (c) are also consistent with each other although they do not seem to converge.
Such an adequacy can be interpreted as the consequence of the asymptotic expansion~\eqref{eq:LowEnergyHamiltonH} of $h^+_0(\eta)$.
Furthermore, Fig.~\ref{fig:fit-pp-delta-fct}(c) shows that the $\Delta_0$-based extrapolation provides reliable results as long as enough experimental points are known in the Coulomb range.
Since this condition is generally satisfied with heavy and moderately nuclei, the direct fitting of $\Delta_0(E)$ is useful for the analysis of weakly bound states.
For instance, in case (c), bound states would correspond to negative-energy zeroes of $\Delta^\fit_0(E)$.

\section{Conclusions}
In this paper, the effective-range function method for charged particles has been studied as well as recently proposed variants~\cite{Sparenberg2017, Blokhintsev2017a, Blokhintsev2017b}.
The usual ERF method, due to Landau~\cite{Landau1944} and Bethe~\cite{Bethe1949}, allows us to interpolate the experimental phase shifts at low energy with a minimum of fitting parameters.
In this way, positive energy data can be extrapolated to the complex $E$-plane to determine resonances and bound states. \par We have given a detailed proof of the expression of the usual ERF by a novel approach solely involving the properties of the Coulomb wave functions in the $E$-plane.
We have established the connection between the original writing~\eqref{eq:UsualERF} of the ERF and the alternate formulation~\eqref{eq:HamiltonUsualERF} due to Cornille and Martin~\cite{Cornille1962} and Hamilton \textit{et al.}~\cite{Hamilton1973} based on the function $h^+_\ell(\eta)$.
We have also shown that the reduced ERF $\Delta_\ell$~\cite{Sparenberg2017} has special structures at negative energy: an accumulation point of poles and zeroes as well as a branch cut emanating from the principal-valued logarithm.
These structures, also seen in the complementary function $g_\ell(\eta)$, make the reduced ERF $\Delta_\ell$ singular at $E=0$.
We have graphically verified the expected properties of the functions $\Delta_\ell$ and $\Delta^+_\ell$ in the $E$-plane for the well-known proton-proton $^1S_0$ collision.
\par As pointed out in Ref.~\cite{Sparenberg2017}, the function $\Delta_\ell$ is in practice much smaller than $g_\ell(\eta)$ for heavy and moderately heavy nuclei at low energy, because of the prefactor $\pi/(\E^{2\eta\pi}-1)$.
Therefore, the addition of $g_\ell(\eta)$ could bias the heuristic interpolation of the usual ERF $\varkappa_\ell$, leading to an underfitting of the phase-shift-dependent part $\Delta_\ell$~\cite{Orlov2016b, Orlov2017b}.
To avoid this, we propose to interpolate $\Delta_\ell$ by means of Eq.~\eqref{eq:DeltaExpansion} considering $-g_\ell(\eta)$ as the first term of the expansion, in accordance with the usual ERF theory.
\par A potential alternative proposed in Ref.~\cite{Sparenberg2017} is to directly interpolate $\Delta_\ell$ by Padé approximants, being closer to the phase shift.
Caution should be exercised when using this method because the Padé approximants are not expected to faithfully converge to the function $\Delta_\ell$, given its singularities.
\par However, this approach turns out to be heuristically useful in two different cases: either to determine resonances, or to study weakly bound states ($\lesssim 1~\Ryd$), as long as data are known in appropriate energy ranges.
With bound states, the method exploits the noticeable property that $\Delta_\ell$ and $\Delta^+_\ell$ join smoothly together at $E=0$ in the physical sheet, due to the common asymptotic Stirling expansion of $g_\ell(\eta)$ and $h^+_\ell(\eta)$ at this point.
This property allows us to reliably extrapolate data below $1~\Ryd$ to negative energy in $\abs{E}\lesssim 1~\Ryd$ in the physical sheet with expansions of relatively low order.
In practice, obtaining a reliable interpolation on the two ranges $E\ll1~\Ryd$ and $E\gg1~\Ryd$ turns out to be difficult, likely because of the sharp inflection point of $\Delta_\ell$ at $E\simeq 1.1~\Ryd$.
For this reason, it seems preferable to restrict the data points to specific energy ranges when fitting the Padé approximants.
\par Finally, the present study theoretically justifies in which situations the low-energy scattering of charged particles can be directly parametrized in terms of a Taylor or Padé expansion of the $\Delta_\ell$  function, as was empirically found for the ${}^{12}\mathrm{C}+\alpha$ system \cite{Sparenberg2017, Blokhintsev2017a, Blokhintsev2017b}.
For other systems, like proton-proton, the use of the standard ERF is still required, at least to compensate for the lack of experimental data at energies around and below the nuclear Rydberg energy, where the mathematical singularities of the Coulomb functions play an crucial role.
With these guidelines in mind, other systems can be tackled.
\par In the future, we plan to further study the interest of Padé approximants, for either the reduced or the standard effective-range functions, to extend the parametrization studied here for low-energy data up to high energies.
We also plan to expand the present results to other reaction channels and to coupled-channel situations.

\begin{acknowledgments}
The work presented in this paper was supported in part by the the IAP program P7/12 of the Belgian Federal Science Policy Office.
It also received funding from the European Union's Horizon 2020 research and innovation program under Grant Agreement No. 654002.
\end{acknowledgments}

\appendix
\section{Graphing complex functions\label{sec:ComplexPlot}}
Graphical representation of complex-valued functions of one complex variable ($f:\mathbb{C}\rightarrow\mathbb{C}$) are quite challenging.
Although there are many ways to proceed, we have chosen in this paper to use color-coded phase plots, as recommended in Ref.~\cite{Wegert2012}.
This method is more straightforward to implement and provides less ambiguous graphics than 3D plots~\cite{Wegert2012}, with the functions that we consider.
However, for convenience, the complementary 2D plots along the real axis are shown throughout this paper.
\par This graphical method consists in representing the complex argument with a hue on the color wheel.
Indeed, the argument turns out to be more useful than the complex modulus to identify the analytic structure of a function, especially the poles, zeroes and branch cuts.
In practice, poles can be graphically distinguished from zeroes using Cauchy's argument principle~\cite{Wegert2012}.
\begin{figure}[ht]
\inputpgf{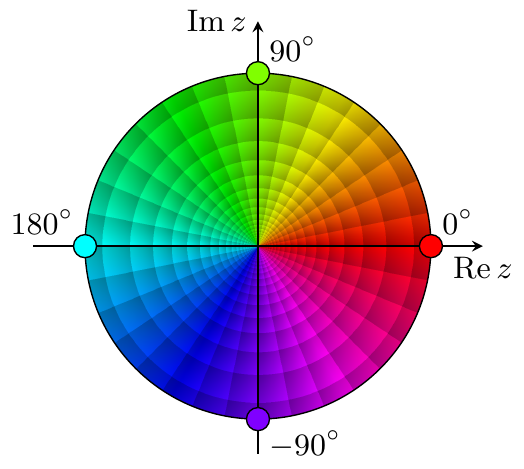}
\caption{\label{fig:identity}(Color online) Identity function in the complex plane with the color code used in this paper.
The complex argument is represented with a hue: red color at $0^\circ$ (for $\mathbb{R}^+$), chartreuse green at $90^\circ$, cyan at $180^\circ$ (for $\mathbb{R}^-$) and violet at $-90^\circ$. The log-polar grid highlights both the modulus and the argument.}
\end{figure}
\par The color map used in this paper is shown in Fig.~\ref{fig:identity} for the identity function $z\mapsto z$.
The analytic structures are also highlighted by an array of contour lines in phase and in modulus forming a logarithmic polar grid.
This logarithmic polar grid allows us to directly visualize the conformality of the mapping $z\mapsto f(z)$ --- and therefore the analyticity of $f(z)$ ---  through the preservation of right angles of the tiles~\cite{Wegert2012}.
In practice, the polar grid is obtained by modulation of the color value $v(z)$ according to the formula  \begin{equation}\label{eq:PolarGrid}
v(z) = v_0 + (1-v_0) \frac{\fpart\big(\frac{N}{2\pi}\ln\abs{z}\big) + \fpart\big(\frac{N}{2\pi}\arg z\big)}{2}  \:,
\end{equation}
where $\fpart{x}$ denotes the fractional part, also known as the sawtooth function, defined by $\fpart{x} = x - \lfloor x\rfloor$ with the floor function $x\mapsto\lfloor x\rfloor$.
We have chosen to set the minimum color value $v_0$ to $60\%$ and the number of angular divisions $N$ to $32$. \par Many other color codes can produce such a polar grid, but Eq.~\eqref{eq:PolarGrid} has the advantage of not requiring too much extra computing effort, and ensuring for all $N$ that the sides of the tiles look as equal as possible, independently of the modulus.

\end{document}